\documentclass[review,3p,times]{elsarticle}
\usepackage{amsmath,hyperref}
\usepackage{lineno}
\usepackage{color}
\usepackage{subfigure}
\usepackage{epsfig}
\usepackage{caption,xcolor,float,graphicx}
\usepackage{graphics}
\usepackage{epstopdf}
\usepackage{float}
\usepackage{booktabs}
\usepackage{multirow}
\usepackage{bm}
\usepackage{amssymb}
\usepackage{booktabs}
\usepackage{dcolumn}
\usepackage{changepage}
\usepackage{chemarrow}
\usepackage{threeparttable}
\usepackage{comment}
\usepackage{adjustbox}
\usepackage{caption}

\biboptions{sort&compress}

\usepackage[T1]{fontenc}
\journal{Applied Mathematical Modelling}

\begin{document}
\captionsetup[figure]{labelfont={bf},name={Fig.},labelsep=period}

\begin{frontmatter}
	
\title{Thermocapillary migration of an odd viscous droplet on a uniformly heated surface: A lattice Boltzmann study}

\author[mymainaddress]{Shan Jiang}
\author[mymainaddress]{He Yan}
\author[mysecondaddress]{Chenxia Xie}
\author[mymainaddress]{Lei Wang\corref{mycorrespondingauthor}} 

\cortext[mycorrespondingauthor]{Corresponding author}
\ead{wangleir1989@126.com}
\address[mymainaddress]{School of Mathematics and Physics, China University of Geosciences, Wuhan 430074, China}
\address[mysecondaddress]{Institute of Cereal \& Oil Science and Technology, Academy of National Food and Strategic Reserves Administration, Beijing 102209, China}

\begin{abstract}
In this study, the thermocapillary actuation behavior of an odd viscous droplet on a uniformly heated surface is numerically investigated using a phase-field-based lattice Boltzmann method. The numerical results reveal that unlike a  conventional viscous droplet that remains stationary on a uniformly heated surface, the presence of odd viscosity converts tangential Marangoni stresses into asymmetric normal stresses along the interface, thereby inducing spontaneous droplet motion. Specifically, when the odd viscosity coefficient $\eta_o/\eta_e$ ($\eta_o$ is odd viscosity, $\eta_e$ is even viscosity) is positive (negative), the droplet migrates toward the right (left). Additionally, due to the enhanced interfacial temperature gradient, the droplet migration velocity consistently increases with the contact angle. Further, it is  observed that the droplet’s migration velocity decreases with an increasing viscosity ratio between the surrounding fluid and the droplet. Finally, as the droplet is placed on an inclined surface, its migration direction and velocity are governed by the interaction between gravity and the odd viscosity-induced force, and in certain cases, the droplet can even climb upward against gravity.

\end{abstract}

\end{frontmatter}
	
\section{Introduction}\label{section1}

Thermocapillary flow is a fluid motion phenomenon driven by a gradient in interfacial tension, which arises from variations in temperature or chemical composition. When this phenomenon is induced by a temperature gradient on a droplet's surface, it typically causes the droplet to move toward the cooler region \cite{Pra Langmuir2008}. The thermocapillary migration of droplets is widely prevalent across many fields, including energy, chemical engineering, and aerospace mechanics. Specific applications include liquid cooling in microelectronics \cite{Hu Langmuir2005,Yan Micromachines2019}, the preparation of multiple emulsions \cite{Utada Science2005,Atencia Nature2005}, and the fabrication of space alloys under microgravity \cite{Jiang Microgravity2019}. Therefore, it is of great scientific and practical importance to understand the principles governing the thermocapillary migration of droplets.

In recent years, the thermocapillary migration of droplets has attracted considerable research interest. Early research in this field was dominated by theoretical analysis, with the pioneering work of Young et al. \cite{Young JFM1959} establishing its theoretical foundation. They derived the first analytical solution for the terminal velocity of a spherical droplet in an infinite ambient fluid. Since then, numerous experiments have been conducted to investigate the thermocapillary migration \cite{Kamotani JFM2000,Kang JFM2019}. Among the earliest experiments on a solid surface were those conducted by Brzoska et al. \cite{Brzoska Langmuir1993}, who investigated droplet movement on a substrate with a thermal gradient. 
Later studies systematically analyzed how migration is affected by parameters such as droplet volume, liquid properties, and temperature gradient \cite{Dai Langmuir2016,Baumgartner PNAS2022}.
These works have been very useful for understanding thermocapillary migration. However, due to the small spatial and temporal scales involved in such migration, precise experimental measurements of local temperature and velocity fields remain a challenging task \cite{Karbalaei Micromachines2016}.

With the development of computer science and numerical methods, numerical simulation has become a powerful tool for studying the thermocapillary migration of droplets. Many scholars have utilized computational fluid dynamics (CFD) methods to research this phenomenon \cite{Bekezhanova AMM2018,Li JAP2024}. For instance, an analysis by Das et al. \cite{Das POF2018} on surfactant-laden droplets in non-isothermal Poiseuille flow found that surface dilatational viscosity significantly retards migration. Wu employed a front-tracking method to investigate the thermocapillary migration of a droplet in a vertical temperature gradient controlled by thermal radiation, demonstrating that the steady migration velocity decreases as the Marangoni number increases \cite{Wu POF2022}. By using the same numerical method, Nguyen et al. \cite{Nguyen EJM2024} simulated the thermal migration of a compound droplet on a substrate. The results showed that the size of the inner core is a key factor in determining the entire droplet's migration direction. Furthermore, the migration of sessile droplets was investigated by Wang et al. \cite{Wang IECR2023} using a three-dimensional level-set method, whose study concluded that the direction of movement is primarily dependent on the contact angle. In a numerical study of droplet migration on an oleophilic track, Kalichetty et al. \cite{Kalichetty IJHMT2023} determined that an optimal track width exists, although restricting the droplet's transverse spread can enhance its migration velocity. More recently, the behavior of self-rewetting droplets on an inclined surface was simulated by Yan et al. \cite{Yan ATE2025} to analyze the effects of the Marangoni number and viscosity ratio on droplet deformation and internal vortex structures.

The above studies have advanced the understanding of thermocapillary migration. However, it is noted that the existing body of work is primarily confined to conventional fluids, whose deviatoric stress tensor is entirely symmetric, meaning that viscous effects are purely dissipative in nature \cite{Landau FM1987,Avron JSP1998,Fruchart AR2023}. In recent years, a special class of fluids known as odd viscous fluids has become a research frontier in the field of soft matter physics \cite{Souslov PRL2019,Markovich PRL2021,de Nature2024}. Unlike conventional fluids, the deviatoric stress tensor in odd viscous fluids contains an antisymmetric part characterized by a non-dissipative coefficient known as odd viscosity \cite{Avron JSP1998,Fruchart AR2023}. Therefore, for an odd viscous fluid, the deviatoric stress tensor is decomposed into a symmetric even component $\tau_e$ and an antisymmetric odd component $\tau_o$, i.e., $\tau = \tau_e + \tau_o$ \cite{Avron JSP1998,Banerjee Nature2017,Fruchart AR2023}. Until now, only few scholars have shifted their focus to the dynamics of odd viscous droplets. These limited works have confirmed that odd viscous droplets exhibit novel properties, such as asymmetric bouncing and rolling upon impact with a solid surface \cite{Franca 2025}. As for the thermocapillary migration, the pioneering work by Aggarwal et al. \cite{Aggarwal PRL2023} revealed that odd viscosity can induce a net droplet migration by breaking the symmetry of the internal flow field on a uniformly heated horizontal substrate. However, their research scenario was strictly limited to a horizontal substrate and did not consider other key factors such as gravity, surface wettability, or the viscosity ratio.

Framed in this general background, this paper aims to numerically investigate the thermocapillary migration of odd viscous droplets on a uniformly heated surface, with a focus on the effects of the odd viscosity coefficient, surface wettability, viscosity ratio, and the inclination angle of the substrate. For the numerical method, we adopt the phase-field-based lattice Boltzmann (LB) method in this work. Compared with conventional CFD methods, such as the level-set method \cite{Sussman CP1998} and volume of fluid method \cite{Hirt JCP1981}, the LB method features a simple algorithm and is particularly effective for handling complex interfaces in multiphase flows \cite{Graaf Langmuir2006,Huang CF2022}. These advantages have made it a powerful tool for simulating complex fluid dynamics phenomena \cite{T.K SCS2017,Xiong 2025}, with successful applications in thermocapillary flows \cite{Wang PRE2023,Yan ATE2025}. The remainder of this paper is organized as follows. Section \ref{section2} describes the phase-field method employed in this study. Section \ref{section3} introduces the LB method for the temperature field. Section \ref{section4} provides a detailed description of the problem model studied in this paper. In Section \ref{section5}, the numerical method employed is validated. Section \ref{section6} presents and discusses the numerical results. Finally, Section \ref{section7} provides brief conclusions.

\section{Governing equations}\label{section2}

To simulate the thermocapillary migration of odd viscous droplets, this study employs the phase-field method, a widely used diffusive interface approach for multiphase flows. 
As an interface capturing method, it is characterized by the use of the order parameter to describe the phase interface \cite{T.K SCS2017}.
Usually, the most widely adopted interface capturing equation is the so-called Cahn-Hilliard equation \cite{Chiu JCP2011,Hua JCP2011}. 
However, the traditional LB models for solving the Cahn-Hilliard equation involve a non-local collision process \cite{Zhang AMM2022}, which in turn introduces additional difficulties in numerical simulations. In this context, the present study adopts the second-order conservative Allen-Cahn equation, which is expressed as \cite{Wang PRE2023}

\begin{equation}
	\label{eq1}
	\frac{\partial \phi}{\partial t} + \nabla \cdot (\phi \mathbf{u}) = \nabla \cdot \left[ M (\nabla \phi -  \frac{1 - \phi^{2}}{\sqrt{2W}} \mathbf{n}) \right],	
\end{equation}
where $\phi$ is the order parameter, $\mathbf{u} = (u, v)$ is the velocity, $M$ is the mobility, $W$ is the interface thickness, and $\mathbf{n} = \nabla \phi / |\nabla \phi|$ is the unit vector normal to the interface.
In this work, we use $\phi_A = 1$ for the droplet and $\phi_B = -1$ for the external fluid, where the interface between the two fluids can be captured by $\phi = 0$.

In addition to the interface-capturing equation, the governing equations for the two-phase flows are also required. Assuming the two fluids are immiscible and incompressible, the Navier-Stokes equations can be written as \cite{Franca 2025}
\begin{subequations}
	\label{eq2}
	\begin{align}
	&\nabla \cdot \mathbf{u} = 0,\\
	\frac{\partial(\rho \mathbf{u})}{\partial t} + \nabla \cdot (\rho \mathbf{u} \mathbf{u}) = -&\nabla p + \nabla \cdot \bm{\tau}_e + \nabla \cdot \bm{\tau}_o
	+\bm{F}_s + \bm{G}.
	\end{align}
\end{subequations}
Here, $\rho$ is the fluid density, $p$ is the pressure, $\bm{G}$ is the body force, $\bm{\tau}_e$ denotes the even stress tensor and $\bm{\tau}_o$ is the odd stress tensor originating from the antisymmetric part of the viscosity tensor \cite{Avron JSP1998,Banerjee Nature2017}. In such a case, the total deviatoric stress tensor in the odd viscous fluid can be expressed as $\bm{\tau} = \bm{\tau}_e + \bm{\tau}_o$, where $\bm{\tau}_e$  and $\bm{\tau}_o$ are given by \cite{Franca 2025}
\begin{equation}
	\label{eq3}
	\begin{aligned}
	&\bm{\tau}_e = \eta_e 
	\begin{bmatrix}
	2\frac{\partial u}{\partial x} & \frac{\partial u}{\partial y} + \frac{\partial v}{\partial x} \\
	\frac{\partial u}{\partial y} + \frac{\partial v}{\partial x} & 2\frac{\partial v}{\partial y}
	\end{bmatrix},
	&&
	\bm{\tau}_o = \eta_o  
	\begin{bmatrix}
	-\left(\frac{\partial u}{\partial y} + \frac{\partial v}{\partial x}\right) & \frac{\partial u}{\partial x} - \frac{\partial v}{\partial y} \\
	\frac{\partial u}{\partial x} - \frac{\partial v}{\partial y} & \frac{\partial u}{\partial y} + \frac{\partial v}{\partial x}
	\end{bmatrix},
	\end{aligned}
\end{equation}
in which  $\eta_e$  represents the even viscosity and $\eta_o$ denotes the odd viscosity. Based on Eq. (\ref{eq3}), one can clearly find that the even stress tensor  $\bm{\tau}_e$ shares the same formulation with the conventional fluid \cite{Landau FM1987}, while the odd stress tensor  $\bm{\tau}_o$ just belongs to the odd viscous fluids. 
Apart from the above mentioned physical parameters, the surface tension force  $\bm{F}_s$ is defined by the following double-well form  \cite{Wang PRE2023}
\begin{equation}
	\label{eq4}
	\bm{F}_s  = \frac{3\sqrt{2}}{4} \left[|\nabla \phi|^2 \nabla \gamma - \nabla \gamma \cdot (\nabla \phi \nabla \phi) + \frac{\gamma}{W^2} \mu_\phi \nabla \phi\right].
\end{equation}
Here, $\gamma$ is the surface tension  and $\mu_\phi$ is the chemical potential, which can be expressed as \cite{Wang PRE2023}
\begin{equation}
	\label{eq5}
    \mu_\phi = (\phi - \phi_A)(\phi - \phi_B)[\phi - 0.5(\phi_A + \phi_B)] -  W^2\nabla^2 \phi. 
\end{equation}
Furthermore, in order to incorporate the thermocapillary effects, an equation of state must be defined to describe how the surface tension $\gamma$ depends on temperature $T$. In this study, we focus on a linear relationship between $\gamma$ and $T$ unless stated otherwise, which is given by \cite{C.Ma IJMF2011}
\begin{equation}
	\label{eq6}
	\gamma(T) = \gamma(T_{\mathrm{ref}}) - \gamma_T (T - T_{\mathrm{ref}}),
\end{equation}
where $T_{\mathrm{ref}}$ is the reference temperature, and $\gamma_T$ is the rate of variation of the surface tension with the temperature.

In the thermocapillary flow, a crucial element in understanding thermal multiphase flow is to comprehend the time evolution of the temperature field. By neglecting viscous dissipation, the governing equation for the temperature field in an incompressible flow is expressed by \cite{Wang PRE2023}
\begin{equation}
	\label{eq7}
	\rho c_p \left( \frac{\partial T}{\partial t} + \mathbf{u} \cdot \nabla T \right) = \nabla \cdot \lambda \nabla T,
\end{equation}
in which $\lambda$ is the thermal conductivity, $c_p$ is the specific-heat capacity.

\section{Methodology}\label{section3}
Based on the collision operator, the LB method is typically classified into single-relaxation-time (BGK) \cite{Qian EL1992}, two-relaxation-time, and multiple-relaxation-time (MRT) models \cite{I.G CCP2008, Luo PRE2000}. This work employs the BGK model due to its simplicity and efficiency. Specifically, our framework consists of three distinct LB models for the phase field, the velocity field \cite{Wang PRE2016}, and the temperature field \cite{Wang PRE2023}. Additionally, a wetting boundary condition is implemented to accurately describe fluid-surface interactions \cite{Ding PRE2007}.

\subsection{LB model for Allen-Cahn equation}
For the phase field, the LB equation for the Allen-Cahn equation with the BGK collision operator can be expressed as \cite{Wang PRE2016}
\begin{equation}
	\label{eq8}
	f_i(\mathbf{x} + \mathbf{c}_i \delta_t, t + \delta_t) - f_i(\mathbf{x},t) = 
	-\frac{1}{\tau_f} \Bigl[ f_i(\mathbf{x},t) - f_i^{\text{eq}}(\mathbf{x},t) \Bigr] 
	+ \delta_t F_i(\mathbf{x},t),
\end{equation}
in which $f_i(\mathbf{x},t)$ is the discrete distribution function of the order parameter $\phi$ at position $\mathbf{x}$ and time $t$, and $f_i^{\text{eq}}(\mathbf{x},t)$ is the local equilibrium distribution function defined by \cite{Wang PRE2016}
\begin{equation}
	\label{eq9}
	f_i^{\text{eq}} = \omega_i \phi \left( 1 + \frac{\mathbf{c}_i \cdot \mathbf{u}}{c_s^2} \right).
\end{equation}
In this work, we adopt the standard two-dimensional nine-velocity lattice model \cite{T.K SCS2017}, where the weight coefficient  $\omega_i$ and the discrete velocity $\mathbf{c}_i$ are defined as 
\begin{equation}
	\label{eq10}
	\omega_i = 
	\begin{cases}
		4/9, & i=0,\\
		1/9, & i=1,\cdots,4,\\
		1/36,& i=5,\cdots,8,
	\end{cases}
\end{equation}
\begin{equation}
	\label{eq11}
	\mathbf{c}_i=
	\begin{cases}
		(0,0)c, & i=0,\\
		(\cos[(i - 1)\pi/2], \sin[(i - 1)\pi/2])c, & i=1,\cdots,4,\\
		\sqrt{2}(\cos[(i - 5)\pi/2+\pi/4], \sin[(i - 5)\pi/2+\pi/4])c, & i=5,\cdots,8,
	\end{cases}
\end{equation}
where $c=\delta_x/\delta_t$ is the lattice speed and the sound speed $c_s$ is defined as $c_s = c/\sqrt3$.
To recover the Allen-Cahn equation precisely  with the multiscale analysis, the discrete source term $F_i(\mathbf{x}, t)$ in Eq. (\ref{eq8}) should be designed as
\begin{equation}
	\label{eq12}
	F_i = \left(1 - \frac{1}{2\tau_f}\right) \frac{\omega_i \mathbf{c}_i \cdot \left[ \partial_t (\phi \mathbf{u}) + c_s^2 \frac{1 - \phi^{2}}{\sqrt{2W}} \mathbf{n} \right]}{c_s^2},
\end{equation}
and the macroscopic value $\phi$ is determined by  \cite{Wang PRE2016}
\begin{equation}
	\label{eq13}
	\phi(\mathbf{x}, t)=\sum_{i} f_{i}(\mathbf{x}, t).
\end{equation}
By using the Chapman-Enskog analysis, the Allen-Cahn equation can be recovered accurately from the LB equation given by Eq. (\ref{eq8}) with the mobility $M = c_s^2 \left( \tau_f - 0.5 \right) \delta t$.

\subsection{LB model for Navier-Stokes equations}
For the flow field, we need another LB model to solve the Navier-Stokes equations. Based on previous work, the corresponding LB equation is given by \cite{Wang PRE2023}
\begin{equation}
	\label{eq14}
	g_i(\mathbf{x} + \mathbf{c}_i \delta_t, t + \delta_t) - g_i(\mathbf{x}, t) = - \frac{1}{\tau_g} \left[ g_i(\mathbf{x}, t) - g_i^{\text{eq}}(\mathbf{x}, t) \right] + \delta_t G_i(\mathbf{x}, t),
\end{equation}
in which $g_i(\mathbf{x}, t)$ is the discrete velocity distribution function, $g_i^{\text{eq}}(\mathbf{x}, t)$ is the local equilibrium distribution function designed by \cite{Wang PRE2023}
\begin{equation}
	\label{eq15}
	g_i^{\mathrm{eq}} = 
	\begin{cases}
		\displaystyle
		\frac{p}{c_s^2}(\omega_i - 1) + \rho s_i(\mathbf{u}), & i = 0, \\
		\displaystyle
		\frac{p}{c_s^2}\omega_i + \rho s_i(\mathbf{u}), & i \neq 0,
	\end{cases}
\end{equation}
with
\begin{equation}
	\label{eq16}
	s_i(\mathbf{u}) = \omega_i \left[ \frac{\mathbf{c}_i \cdot \mathbf{u}}{c_s^2} + \frac{(\mathbf{c}_i \cdot \mathbf{u})^2}{2 c_s^4} - \frac{\mathbf{u} \cdot \mathbf{u}}{2 c_s^2} \right].
\end{equation}
Different from previous LB models, the discrete source term $G_i(\mathbf{x}, t)$ is expressed as 
\begin{equation}
	\label{eq17}
	G_i = \left(1 - \frac{1}{2\tau_g} \right)\omega_i \left[ \mathbf{u} \cdot \nabla \rho + \frac{\mathbf{c}_i \cdot \mathbf{F}}{c_s^2} + \frac{\mathbf{u} \nabla \rho : \left( \mathbf{c}_i \mathbf{c}_i - c_s^2 \mathbf{I} \right)}{c_s^2} \right],
\end{equation}
where $\mathbf{F} = \mathbf{F_s} + \mathbf{F_o} + \mathbf{G}$ is the total force, and $\mathbf{F_o}$ is provided by the odd viscosity. 
Through the Chapman-Enskog analysis, the incompressible Navier-Stokes equations can be recovered to second-order accuracy with $\mu = \rho c_s^2 \left( \tau_g - \frac{1}{2} \right) \delta t$, and the macroscopic quantities can be calculated by \cite{Wang PRE2023}
\begin{equation}
	\label{eq18}
	\rho \mathbf{u} = \sum_{i} \mathbf{c}_i g_i + 0.5 \delta t \mathbf{F},
\end{equation}
\begin{equation}
	\label{eq19}
	p = \frac{c_s^2}{(1 - \omega_0)} \left[ \sum_{i \neq 0} g_i + \frac{\delta t}{2} \mathbf{u} \cdot \nabla \rho + \rho s_0(\mathbf{u}) \right].
\end{equation}

\subsection{LB model for temperature equation}
In order to solve the temperature field, the improved  thermal LB model proposed by Wang et al. \cite{Wang PRE2023}  is adopted. In contrast to previous LB models, the most distinct feature of this model lies in its ability to incorporate the influence of the heat capacity, and it also exhibits a better numerical accuracy and stability. The LB equation of this improved model can be given by  
\begin{equation}
	\label{eq20}
	\rho c_p h_i \left( \mathbf{x} + \mathbf{c}_i \delta t, t + \delta t \right) - h_i(\mathbf{x}, t) = \left( \rho c_p - 1 \right) h_i \left( \mathbf{x} + \mathbf{c}_i \delta t, t \right) - \frac{1}{\tau_h} \left[ h_i(\mathbf{x}, t) - h_i^{eq}(\mathbf{x}, t) \right] + \delta tH_i(\mathbf{x}, t),
\end{equation}
in which $h_i(\mathbf{x}, t)$ is the temperature distribution function, $\tau_h$ is the dimensionless relaxation time defined as $\tau_h = 0.5 + \lambda/c_s^2 \delta t$,  $h_i^{\text{eq}}(\mathbf{x}, t)$ is the equilibrium distribution function given by \cite{Wang PRE2023} 
\begin{equation}
	\label{eq21}
	h_i^{eq} = \omega_i T,
\end{equation}    
and the discrete source term $H_i(\mathbf{x}, t)$ is defined as \cite{Wang PRE2023}       
\begin{equation}
	\label{eq22}
	H_i = - \omega_i \rho c_p \mathbf{u} \cdot \nabla T.
\end{equation}
Further, the macroscopic temperature  $T$ is determined by the distribution functions as \cite{Wang PRE2023} 
\begin{equation}
	\label{eq23}
	T = \sum_{i} h_i.
\end{equation}

\subsection{Wetting boundary condition}
When considering the effect of surface wettability in the simulation, it is essential to implement a wetting boundary condition capable of prescribing a specified contact angle $\theta$.
To this end, this work adopts the geometric formulation proposed by Ding and Spelt \cite{Ding PRE2007}, which is expressed as
\begin{equation}
	\label{eq24}
	\mathbf{n}_w \cdot \nabla \phi = - \tan \left( \frac{\pi}{2} - \theta \right) \left| \mathbf{n}_\tau \cdot \nabla \phi \right|,
\end{equation}
where $\mathbf{n}_w$ and $\mathbf{n}_\tau$ are the unit vectors normal and tangential to the solid surface, respectively. 
For numerical implementation, the normal component and tangential component of $\nabla \phi$ are discretized using a second-order scheme as \cite{Ding PRE2007}
\begin{subequations}
	\label{eq25}
	\begin{align}
	\mathbf{n}_w \cdot \nabla \phi &= \frac{\phi_{x,1} - \phi_{x,0}}{\Delta x},\\
	\mathbf{n}_\tau \cdot \nabla \phi = \frac{\partial \phi_{x,1/2}}{\partial x} &= 1.5 \frac{\partial \phi_{x,1}}{\partial x} - 0.5 \frac{\partial \phi_{x,2}}{\partial x},
	\end{align}
\end{subequations}
In this case, the wetting boundary condition could be easily implemented by using Eq. (\ref{eq25}).

\section{Problem statement}\label{section4}

\begin{figure}[H]
	\centering
	\includegraphics[width=0.5\textwidth]{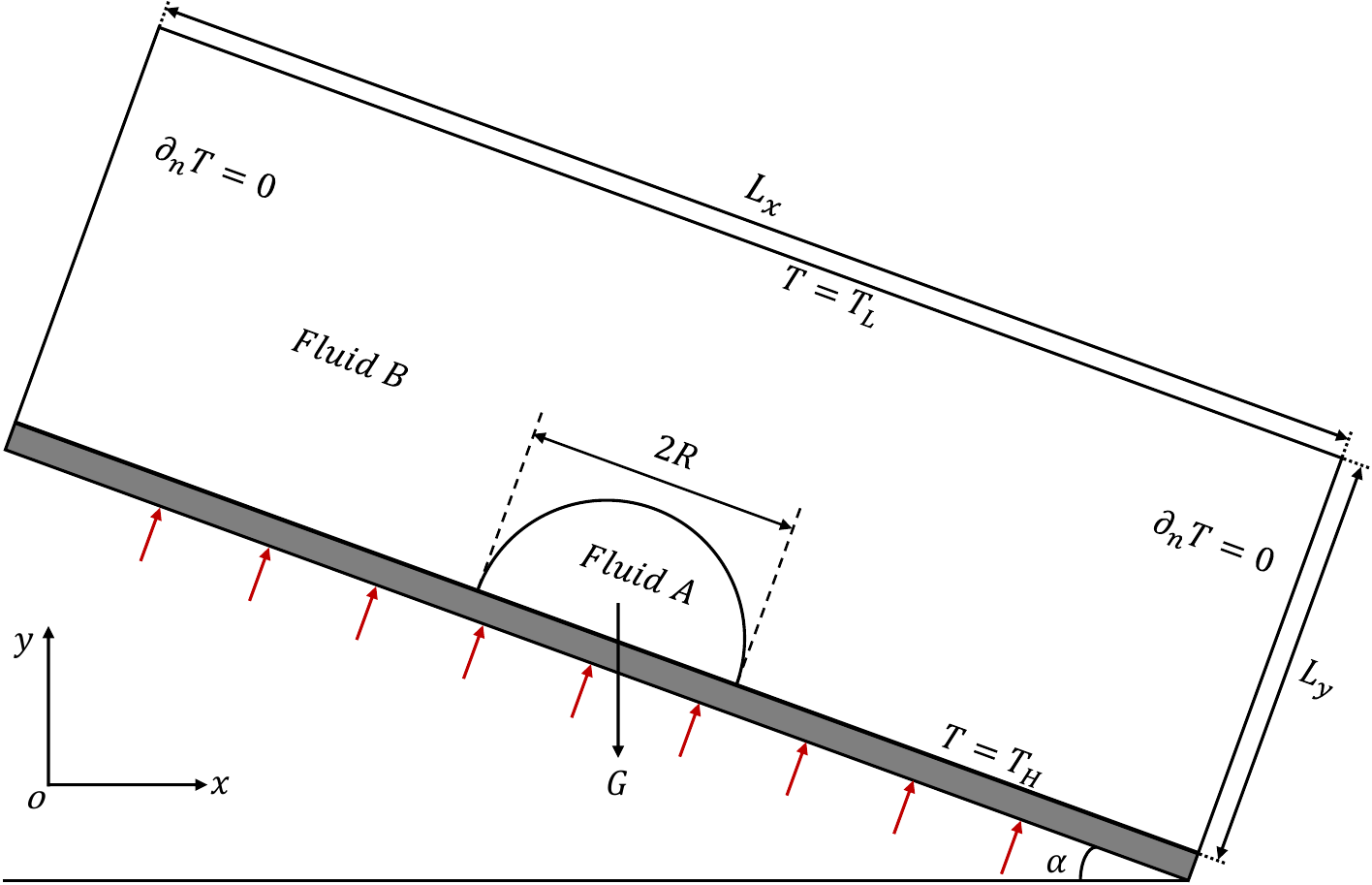}
	\caption{Schematic of a droplet on a uniformly heated inclined solid substrate, where the bottom wall is maintained at a constant temperature $T_H$, the top wall is maintained at a constant temperature $T_L$, and the side walls are adiabatic.}
	\label{fig1}
\end{figure}

In this section, we present the conditions and basic assumptions of this study.
As shown in Fig. \ref{fig1}, a hemispherical droplet with a radius of $R = 50$ is initially placed at rest on a smooth substrate, centered at $(x_0, y_0) = (400, 0)$. The substrate is inclined at an angle $\alpha$ with the negative $x$-axis, and the simulation is conducted within a rectangular computational domain of size $L_x\times L_y = 800\times160$.
All boundaries of the computational domain are considered stationary walls with no-slip conditions. For the thermal boundary conditions, 
the bottom and top boundaries are maintained at constant high ($T_H$) and low ($T_L$) temperatures,
while the left and right boundaries are adiabatic, i.e., $\partial_n T = 0$.
Following the previous study \cite{Liu JCP2015}, $T_H$ and $T_L$ are set to 80 and 0, respectively. 

The thermocapillary migration of an odd viscous droplet can be characterized by the following dimensionless parameters:
Reynolds number ($\mathrm{Re}$), capillary number ($\mathrm{Ca}$), Marangoni number ($\mathrm{Ma}$), Bond number ($\mathrm{Bo}$), and the odd viscosity coefficient ($\eta$), which are defined as
\begin{equation}
	\label{eq26}
	\mathrm{Re}=\frac{\rho_A U R}{\mu_A}, 
	\mathrm{Ca}=\frac{U \mu_A}{\gamma_{\mathrm{ref}}}, 
	\mathrm{Ma}=\frac{\rho_A c_{pA} U R}{\lambda_A}, 
	\mathrm{Bo}=\frac{\rho_A g R^2}{\gamma_{\mathrm{ref}}},
	\mathrm{\eta}=\frac{\eta_o}{\eta_e},	
\end{equation}
where $U = {\gamma_T G_T R} / \mu_A$ is the system's characteristic velocity, $G_T$ is the vertical temperature gradient, and $\mathrm{g}$ is the gravity.

\section{Numerical validation}\label{section5}
This section aims to validate the accuracy and reliability of the present LB method. To this end, we simulate the thermocapillary migration of a conventional droplet on the solid surface and compare our results with those reported in the work of Liu et al. \cite{Liu JCP2015}.
The simulation parameters are chosen to be identical to the values used by Liu et al. \cite{Liu JCP2015}.

\begin{figure}[H]
	\centering
	\begin{minipage}[c]{0.03\textwidth}
		\centering (a)
	\end{minipage}
	\begin{minipage}[c]{0.95\textwidth} 
		\adjustbox{valign=c}{\includegraphics[width=1\textwidth]{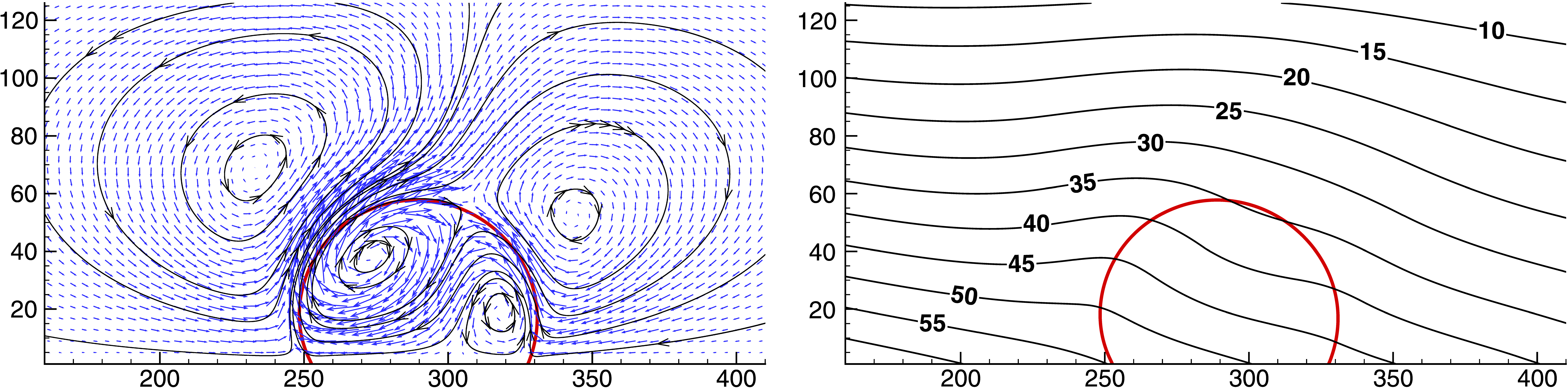}}
	\end{minipage}	
	\vspace{1.0em} 
	\begin{minipage}[c]{0.03\textwidth}
		\centering (b)
	\end{minipage}
	\begin{minipage}[c]{0.95\textwidth}
		\adjustbox{valign=c}{\includegraphics[width=1\textwidth]{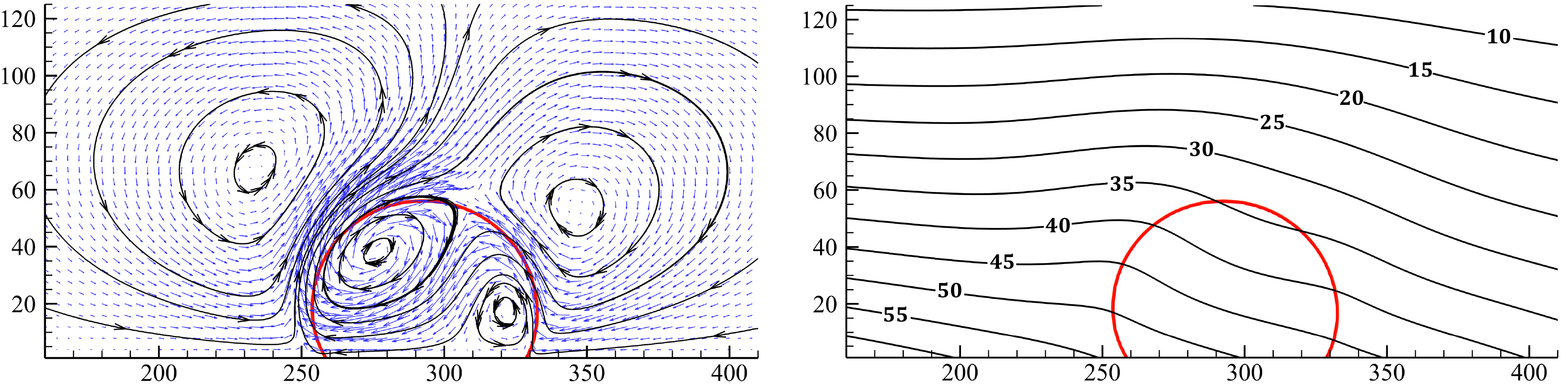}}
	\end{minipage}
	
	\caption{The ﬂow ﬁeld (the left plane) and the temperature ﬁeld (the right plane) surrounding the moving droplet at the contact angle $\theta=120^\circ$, (a) the numerical results reported by Liu et al.\cite{Liu JCP2015}, (b) the present data, in which the red lines are $\phi = 0$, the blue lines with arrows are the velocity vectors, and the black lines with arrows are the streamlines.} 
	\label{fig2}
\end{figure}

\begin{figure}[H]
	\centering
	\includegraphics[width=0.5\textwidth]{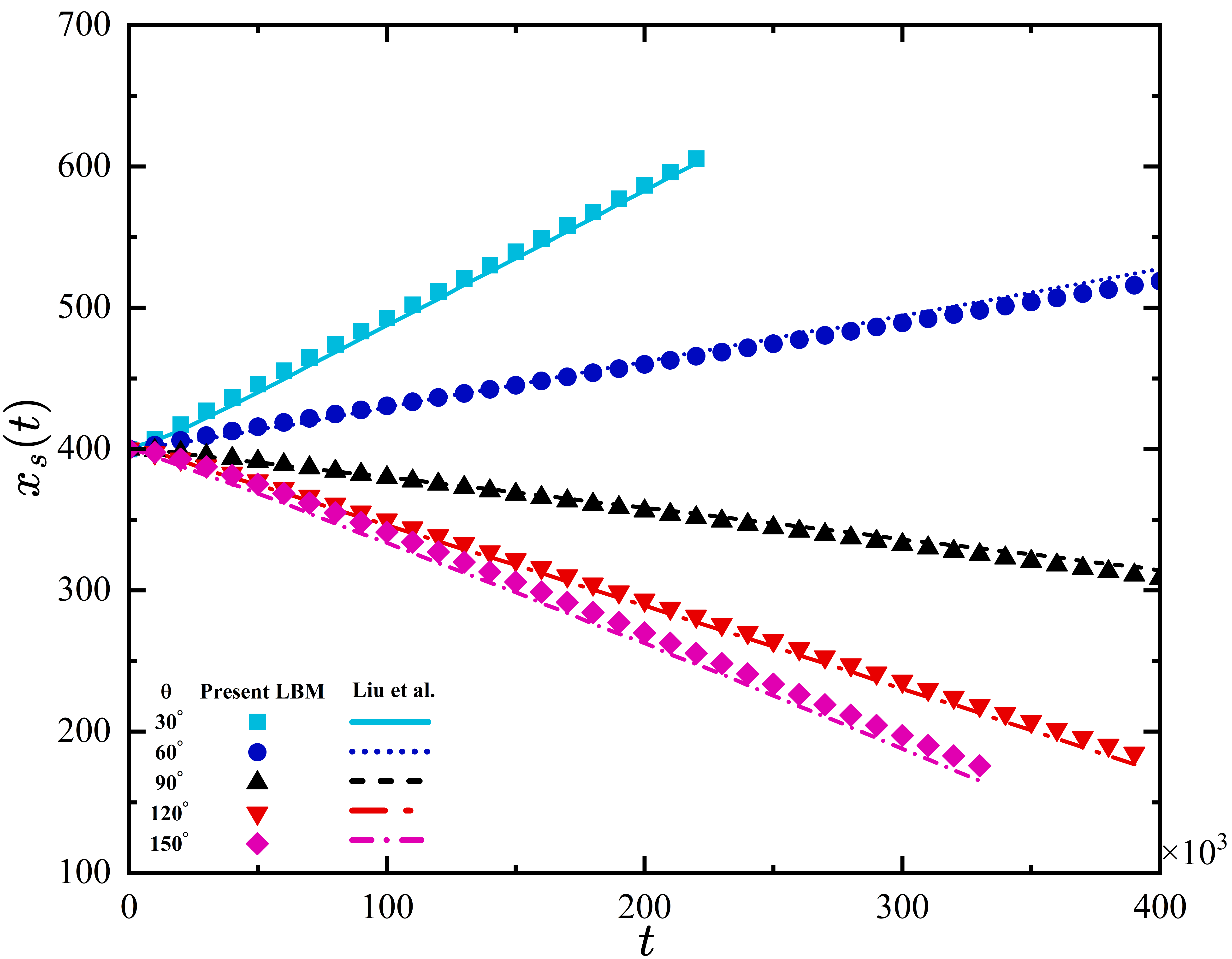}
	\caption{Time evolution of the $x$-coordinate of droplet centroid at various contact angles.}
	\label{fig3}
\end{figure}

Fig. \ref{fig2} shows the streamline distribution (left) and isotherm distribution (right) around the droplet for a contact angle of $\theta=120^\circ$ at the time step of $t=2\times10^5$. 
The results are highly consistent with the findings of Liu et al. \cite{Liu JCP2015}.
Notably, the model successfully reproduces the prominent vortex structures inside the droplet and in the external fluid region.
In addition to providing a qualitative validation of our model, we investigated the temporal evolution of the droplet's centroid during its migration under different contact angles, where the centroid position $x_s (t)$ is expressed by
\begin{equation}
	\label{eq27}
	x_s (t) = \frac{\sum_{\mathbf{x}} x(\mathbf{x}, t)\rho_A (\mathbf{x}, t)}{\sum_{\mathbf{x}} \rho_A (\mathbf{x}, t)}.
\end{equation}
As illustrated in Fig. \ref{fig3}, our numerical results demonstrate excellent agreement with data from previous studies across various contact angles, thus validating the effectiveness of the LB method in simulating dynamics at fluid-solid interfaces.

\section{Results and discussion}\label{section6}
In this section, we aim to simulate the thermocapillary migration of an odd viscous droplet on a solid surface, focusing on the odd viscosity coefficient, surface wettability, viscosity ratio, and inclination angle. 
Following the work of Liu et al. \cite{Liu JCP2015}, the dimensionless parameters are set to ${\lambda_B} /{\lambda_A}=1$, ${c_{pB}}/{c_{pA}}=1$, and the temperature difference $\Delta T$ is fixed at $80$, with a resulting vertical temperature gradient of $G_T = 0.5$ and a characteristic velocity of $U = 0.1$.
In addition, since the effect of gravity is considered in this model, the Bond number is set to $0.01$, and the density ratio is taken to be $\rho_B / \rho_A = 0.8$ \cite{Voc JCS2015}. 
Furthermore, the physical properties of the fluid must align with the order parameter in a two-phase system, and the alignment is maintained through linear interpolation
\begin{subequations}
	\label{eq28}
	\begin{align}
	\rho &= \frac{\rho_A - \rho_B}{\phi_A - \phi_B} \left( \phi - \phi_B \right) + \rho_B, & 
	\mu &= \frac{\mu_A - \mu_B}{\phi_A - \phi_B} \left( \phi - \phi_B \right) + \mu_B, \\
	\lambda &= \frac{\lambda_A - \lambda_B}{\phi_A - \phi_B} \left( \phi - \phi_B \right) + \lambda_B, & 
	c_p &= \frac{c_{pA} - c_{pB}}{\phi_A - \phi_B} \left( \phi - \phi_B \right) + c_{pB}.
	\end{align}
\end{subequations}

\begin{figure}[H]
	\centering
	\begin{minipage}[c]{0.03\textwidth}
		\centering (a)
	\end{minipage}
	\begin{minipage}[c]{0.95\textwidth} 
		\adjustbox{valign=c}{\includegraphics[width=1\textwidth]{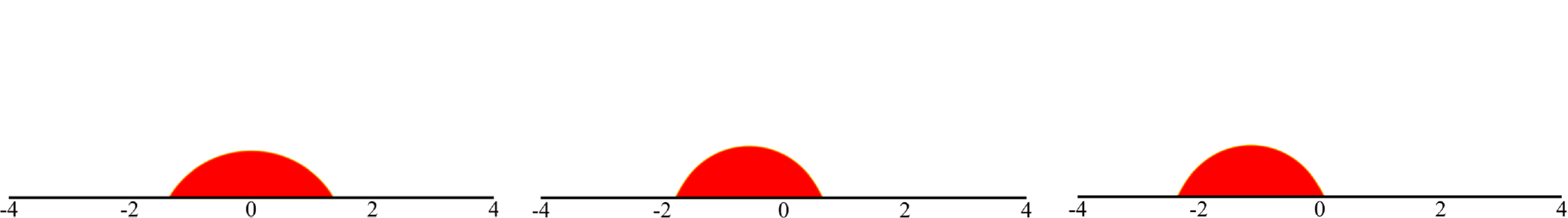}}
	\end{minipage}	
	\vspace{0.0em} 
	\begin{minipage}[c]{0.03\textwidth}
		\centering (b)
	\end{minipage}
	\begin{minipage}[c]{0.95\textwidth}
		\adjustbox{valign=c}{\includegraphics[width=1\textwidth]{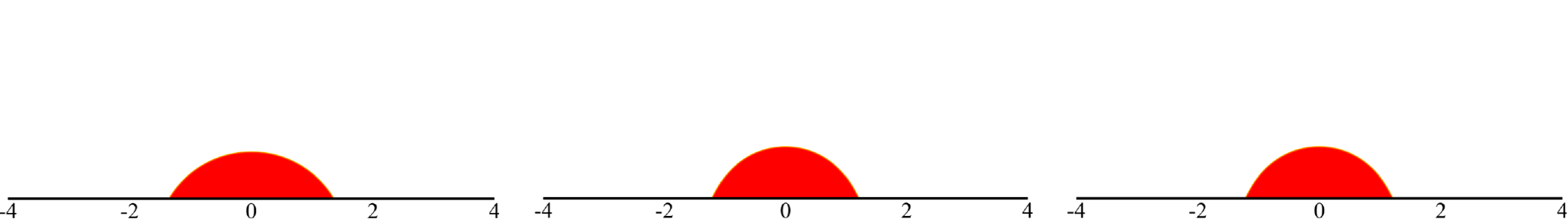}}
	\end{minipage}
	\vspace{0.0em} 
	\begin{minipage}[c]{0.03\textwidth}
		\centering (c)
	\end{minipage}
	\begin{minipage}[c]{0.95\textwidth}
		\adjustbox{valign=c}{\includegraphics[width=1\textwidth]{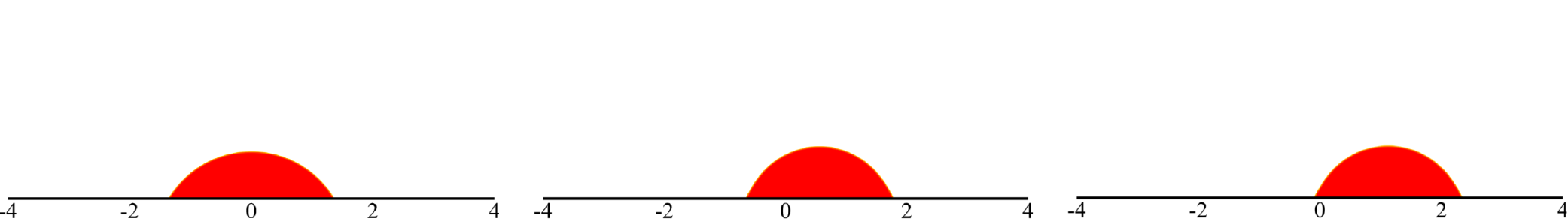}}
	\end{minipage}
	
	\caption{Snapshots of the displacement process for different odd viscosity coefficients: (a) $\eta =-3.0$, (b) $\eta = 0$, (c) $\eta = 3.0$. In each row, images from left to right correspond to dimensionless times $ t^*=0, 4.73, 9.47$, respectively.}
	\label{fig4}
\end{figure}

We first investigate the influence of the odd viscosity coefficient on droplet migration. Initially, a droplet with a contact angle of $60^\circ$ is placed at the center of a flat substrate (at $x = 0$) with an inclination angle of $\alpha = 0^\circ$ for three different values of odd viscosity coefficient ($\eta = -3, 0, 3$). 
Fig. \ref{fig4} presents the temporal evolution of the droplet under each condition, in which the dimensionless time is represented by the Fourier number $t^* = \lambda t / \rho c_p L_y^2$.
Over time, the droplet's dynamical behavior differs significantly depending on the value of $\eta$.
As shown in Fig. \ref{fig4} (b), when $\eta = 0$, the droplet remains stationary throughout the simulation, which is consistent with the behavior of a conventional droplet, where symmetric recirculating flows do not lead to net motion.
In contrast, when $\eta \ne 0$, the droplet exhibits a pronounced directional motion, moving left for $\eta = -3$ and right for $\eta = 3$ [see Figs. \ref{fig4}(a), (c)], which indicates that the direction of motion is governed by the sign of the odd viscosity coefficient.
Particularly, it is interesting to note that Fig. \ref{fig4} also reveals a slight upward bulging of the droplet top over time. The deformation can be attributed to thermocapillary effects induced by bottom heating, which drive fluid from the hotter regions (lower surface tension) to the colder regions (higher surface tension) along the interface, thus promoting the upward accumulation of liquid near the apex of the droplet.

\begin{figure}[H]
	\centering
	\includegraphics[width=0.5\textwidth]{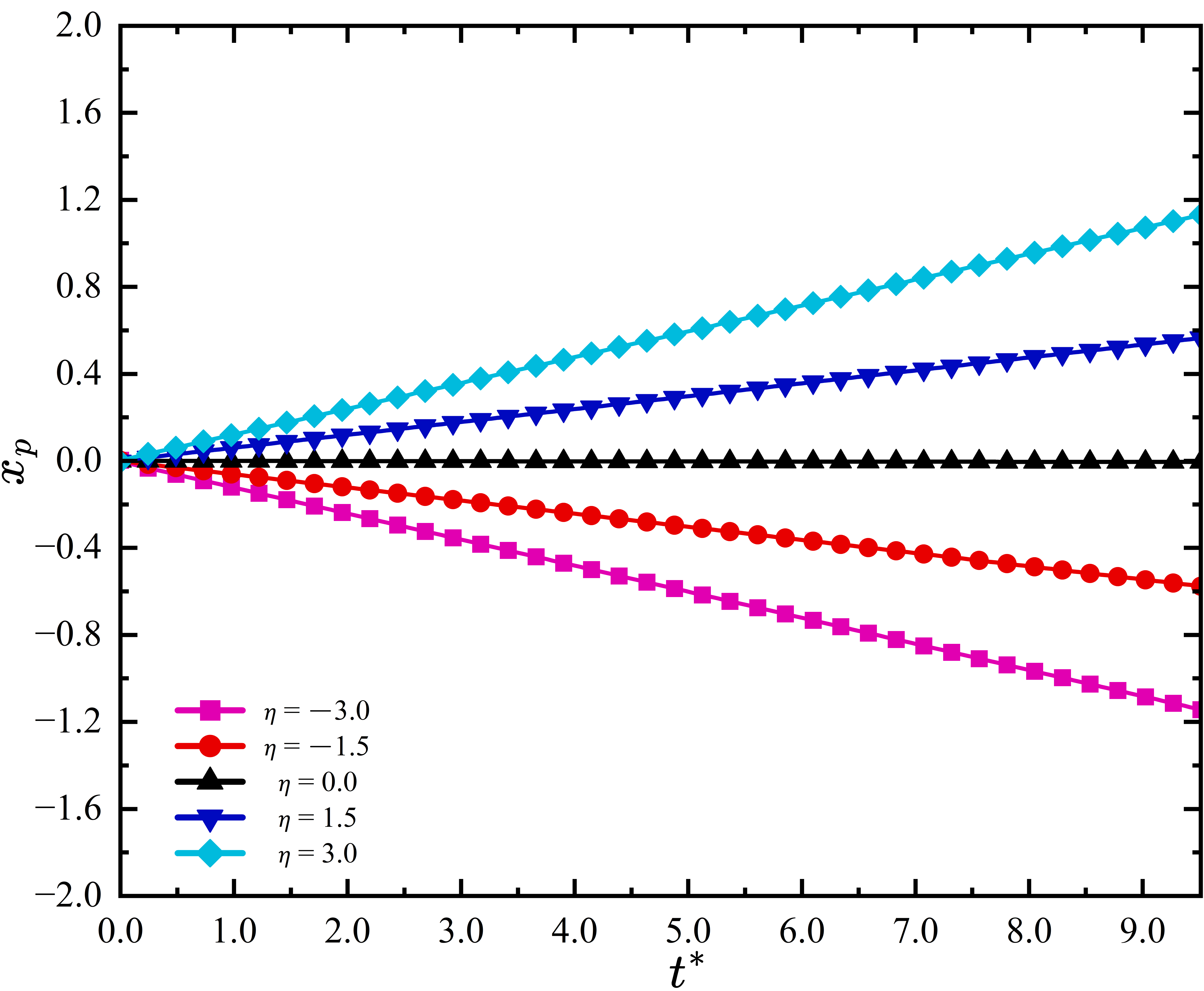}
	\caption{The relationship between centroid position ($x_p$) (normalized by the initial position) and the dimensionless time ($t^*$) at different odd viscosity coefficients.} 
	\label{fig5}
\end{figure}
\begin{figure}[H]
	\centering
	\begin{minipage}[c]{0.03\textwidth}
		\centering (a)
	\end{minipage}
	\begin{minipage}[c]{0.95\textwidth} 
		\adjustbox{valign=c}{\includegraphics[width=1\textwidth]{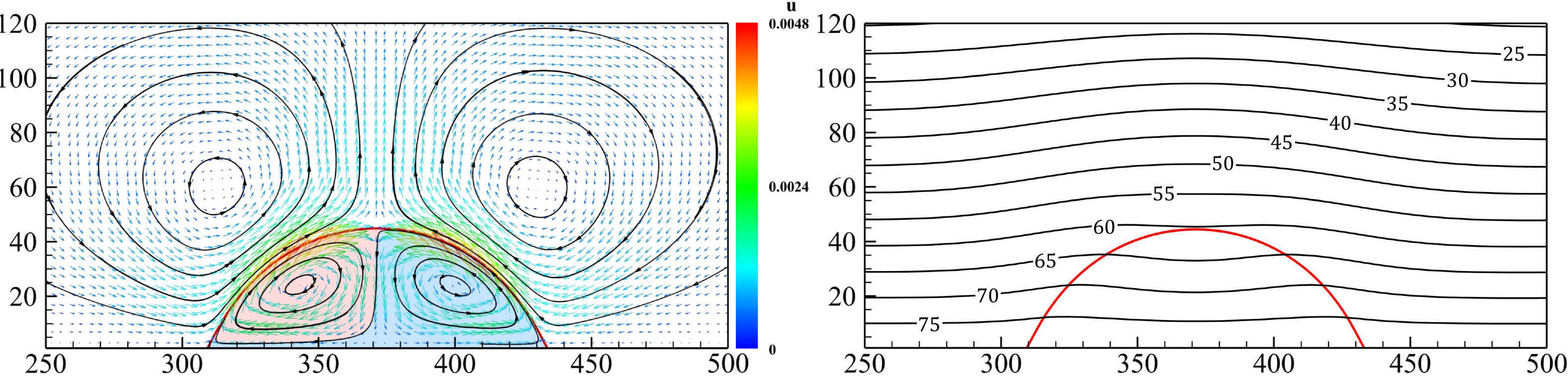}}
	\end{minipage}	
	\vspace{0.0em} 
	\begin{minipage}[c]{0.03\textwidth}
		\centering (b)
	\end{minipage}
	\begin{minipage}[c]{0.95\textwidth}
		\adjustbox{valign=c}{\includegraphics[width=1\textwidth]{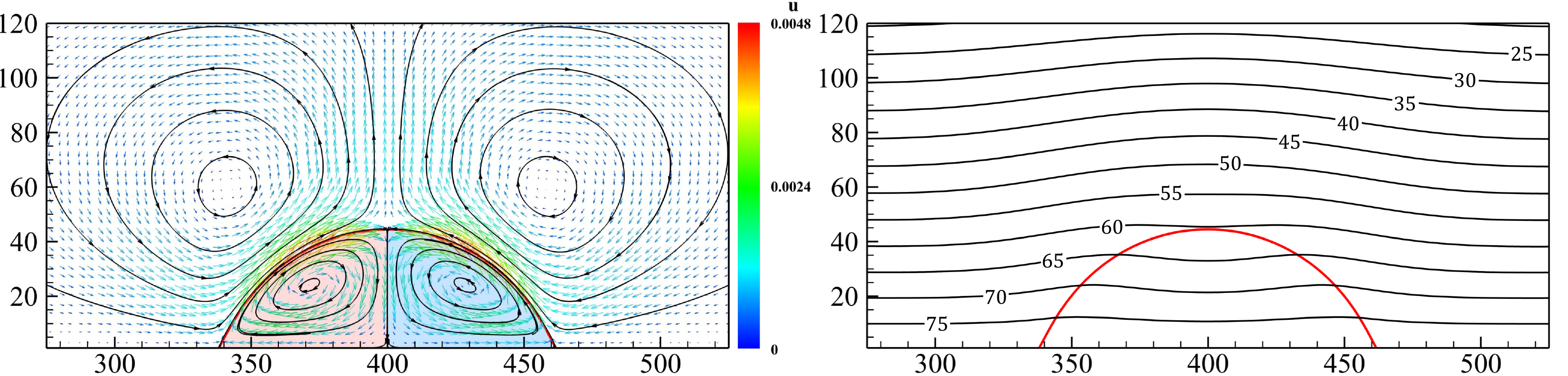}}
	\end{minipage}
	\vspace{0.0em} 
	\begin{minipage}[c]{0.03\textwidth}
		\centering (c)
	\end{minipage}
	\begin{minipage}[c]{0.95\textwidth}
		\adjustbox{valign=c}{\includegraphics[width=1\textwidth]{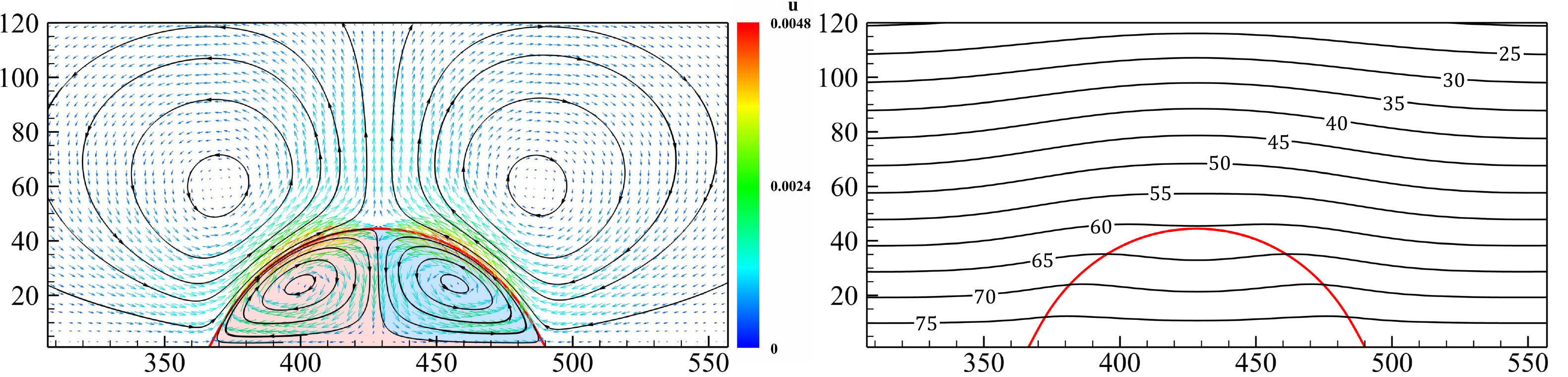}}
	\end{minipage}
	
	\caption{The flow field (the left plane) and the temperature field (the right plane) surrounding the moving droplet at the contact angle $\theta = 60^\circ$ for different odd viscosity coefficients: (a) $\eta$ = -3.0, (b) $\eta$ = 0, (c) $\eta$ = 3.0. The red lines are $\phi = 0$, the colorful lines with arrows are the velocity vectors, and the black lines with arrows are the streamlines.} 
	\label{fig6}
\end{figure}

The relationship between droplet migration and odd viscosity coefficients is  quantified in Fig. \ref{fig5}, which shows the temporal evolution of the droplet centroid position under various values of $\eta$. 
It can be observed that when $\eta = 0$, the centroid position remains nearly constant, whereas for $\eta \ne 0$, the migration velocity of the droplet increases with $|\eta|$.
The result further confirms that odd viscosity serves as the driving mechanism for droplet migration and significantly governs the migration speed.
To have a better understanding of the physical mechanism behind the droplet migration, Fig. \ref{fig6} presents the flow and temperature fields for three representative odd viscosity coefficients ($\eta = -3, 0, 3$).
The temperature field shows that a stable vertical temperature gradient gradually develops from bottom to top within the droplet. 
It is noteworthy that for all values of $\eta$, the temperature distribution remains symmetric with respect to the vertical axis. 
In fact, the odd viscosity term appears only in the momentum equation, while heat transfer is governed by an energy equation that is independent of odd viscosity.
As a result, the temperature field is unaffected by the odd viscosity coefficient.
In sharp contrast, the internal flow field is influenced by odd viscosity.
As shown in the corresponding flow field subfigure of Fig. \ref{fig6}(b), the droplet exhibits two symmetric counter-rotating vortices of equal strength when $\eta = 0$, and therefore remains stationary. 
However, as illustrated in Fig. \ref{fig6}(a) and (c), when $\eta \ne 0$, the two counter-rotating vortices inside the droplet become asymmetric, causing the droplet to migrate along the solid surface.

\begin{figure}[H]
	\centering
	\includegraphics[width=0.5\textwidth]{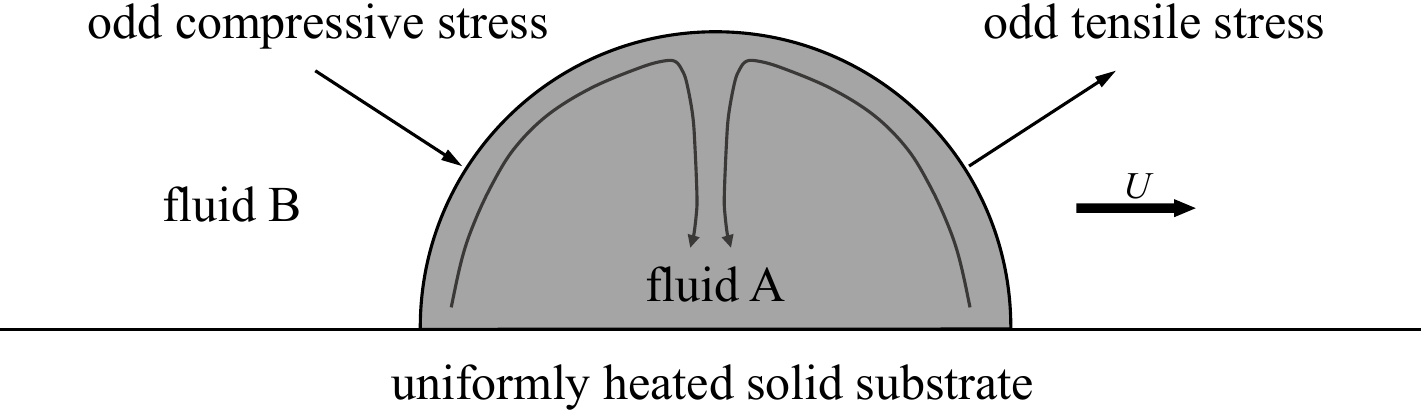}
	\caption{The odd viscosity induces a compressive normal stress along the left segment of the droplet and a tensile stress along the right segment when $\eta_o > 0$, leading to a net imbalance that causes the droplet to  migrate toward the right.} 
	\label{fig7}
\end{figure}

Through an analysis of the Marangoni stresses induced by temperature gradients and the asymmetric constitutive behavior arising from odd viscosity, we can elucidate the mechanism behind the asymmetric response.
Under the boundary condition of uniform heating from the bottom and a constant low temperature at the top, a stable vertical temperature gradient gradually develops within the droplet through thermal conduction, as shown in the temperature field subplots of Fig. \ref{fig6}. Due to curvature variations along the droplet interface, the temperature distribution also varies along the arc length, which results in a surface tension gradient that further induces Marangoni stresses.
On the left segment of the droplet interface, where the temperature gradient is anticlockwise, Marangoni stresses drive fluid away from the region of high temperature (low interfacial tension), thereby generating a clockwise vortex within the left portion of the droplet. In contrast, on the right interface segment, where the temperature gradient is clockwise, the induced flow results in a counterclockwise vortex on the right side.
To better understand the role of odd viscosity in droplet migration, the form of the Cauchy stress tensor $\boldsymbol{\sigma}$ is revisited. In fact, as discussed by Landau and Lifshitz \cite{Landau FM1987}, for conventional incompressible fluids ($\eta_o = 0$), the Cauchy stress tensor $\boldsymbol{\sigma}$ is symmetric and takes the standard (even) viscous form
\begin{equation}
	\label{eq29}
	\boldsymbol{\sigma} = -p \boldsymbol{I} + 2\eta_e \boldsymbol{D},	
\end{equation}
where $\boldsymbol{I}$ is the unit tensor, $D_{ij} = \frac{1}{2} \left( \frac{\partial u_i}{\partial x_j} + \frac{\partial u_j}{\partial x_i} \right)$ is the rate-of-strain tensor.
At the droplet interface, the interfacial stress can be divided into normal and tangential components: the normal component is determined by the pressure, curvature, and external normal stress, while the tangential component is determined by the surface tension gradient and external tangential stress. 
In such a case, the tangential and normal stresses are independent of each other, the symmetry of the system is maintained, and the droplet remains stationary as shown in Fig. \ref{fig4} (b).
However, for odd viscous fluids, the Cauchy stress tensor contains an antisymmetric coupling term and can be expressed as \cite{Avron JSP1998}
\begin{equation}
	\label{eq30}
	\boldsymbol{\sigma} = -p \boldsymbol{I} + 2 \begin{pmatrix}
	\eta_e & -\eta_o \\
	\eta_o & \eta_e
	\end{pmatrix} \boldsymbol{D}.	
\end{equation}
In this context, the tangential stress can induce a normal response, and the normal stress can also influence tangential component, leading to tangential-normal coupling. The coupling can be represented by the rate-of-strain tensor at the interface \cite{Oron RMP1997, Aggarwal PRL2023}

\begin{equation}
	\label{eq31}
	\boldsymbol{t}\boldsymbol{D}\boldsymbol{n} = \frac{1}{2(\eta_e^2 + \eta_o^2)} \left[ \eta_e \underbrace{\left( \frac{\partial \gamma}{\partial s} + \tau_0 \right)}_{\text{the tangential stress-related term}} + \eta_o\underbrace{ (p + 2\kappa\gamma + \Pi) }_{\text{the normal stress-related term}}\right], 
\end{equation}
\begin{equation}
	\label{eq32}
	\boldsymbol{n}\boldsymbol{D}\boldsymbol{n} = \frac{1}{2(\eta_e^2 + \eta_o^2)} \left[ -\eta_o \underbrace{\left( \frac{\partial \gamma}{\partial s} + \tau_0 \right)}_{\text{the tangential stress-related term}} + \eta_e\underbrace{ (p + 2\kappa\gamma + \Pi) }_{\text{the normal stress-related term}}\right],
\end{equation}
where $\boldsymbol{t}$ is the unit vector tangential to the interface, $\boldsymbol{n}$ is the unit outward vector normal to the interface,  $\gamma$ is the (variable) surface tension, $s$ is the arc length along the interface, $\kappa$ is the mean curvature of the interface, $\tau_0$ and $\Pi$ are the tangential and normal components of the prescribed forcing at the interface, respectively.
According to Eqs. (\ref{eq31}) and (\ref{eq32}), one can find that in the odd viscous fluid, two additional stress components arise at the interface: a tangential stress induced by the normal stress, and a normal stress induced by the tangential stress.
Since there is no external fluid applying forces on the interface in this problem (i.e., $ \tau_0 = 0, \Pi = 0$), the normal component of the rate-of-strain tensor $\boldsymbol{D}$ at the interface is given by
\begin{equation}
	\label{eq33}
	\boldsymbol{n}\boldsymbol{D}\boldsymbol{n} = \frac{-\eta_o \frac{\partial \gamma}{\partial s} + \eta_e (p + 2\kappa\gamma)}{2(\eta_e^2 + \eta_o^2)}.
\end{equation}
Therefore, odd viscosity gives rise to a tangential-induced contribution in the normal component of the rate-of-strain tensor, which is proportional to $\eta_o$. More specifically, when $\eta_o > 0$, the coupling results in a compressive stress on the left interface (where ${\partial \gamma}/{\partial s} > 0$), and a tensile stress on the right interface (where ${\partial \gamma}/{\partial s} < 0$), which provides the driving mechanism for the droplet migration [see Fig. \ref{fig7}], consistent with the streamline directions shown in Fig. \ref{fig6} (c).
In summary, through the antisymmetric coupling of odd viscosity in the stress tensor, odd viscosity converts the originally tangential Marangoni stress into an asymmetric normal stress at the interface, which constitutes the intrinsic mechanism underlying the spontaneous directional motion of the droplet.

\begin{figure}[H]
	\centering
	\begin{minipage}[c]{0.03\textwidth}
		\centering(a)
	\end{minipage}
	\begin{minipage}[c]{0.95\textwidth} 
		\adjustbox{valign=c}{\includegraphics[width=1\textwidth]{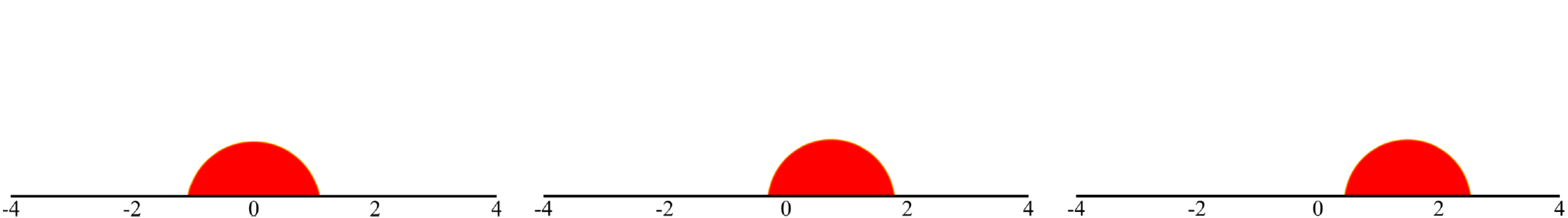}}
	\end{minipage}	
	\begin{minipage}[c]{0.03\textwidth}
		\centering (b)
	\end{minipage}
	\begin{minipage}[c]{0.95\textwidth}
		\adjustbox{valign=c}{\includegraphics[width=1\textwidth]{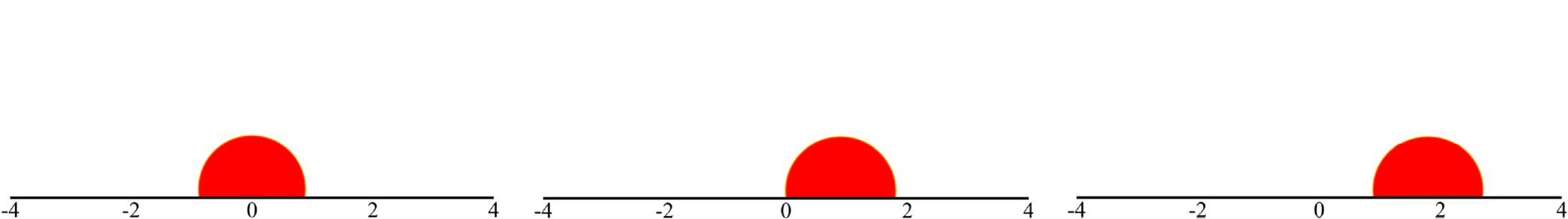}}
	\end{minipage}
	\begin{minipage}[c]{0.03\textwidth}
		\centering (c)
	\end{minipage}
	\begin{minipage}[c]{0.95\textwidth}
		\adjustbox{valign=c}{\includegraphics[width=1\textwidth]{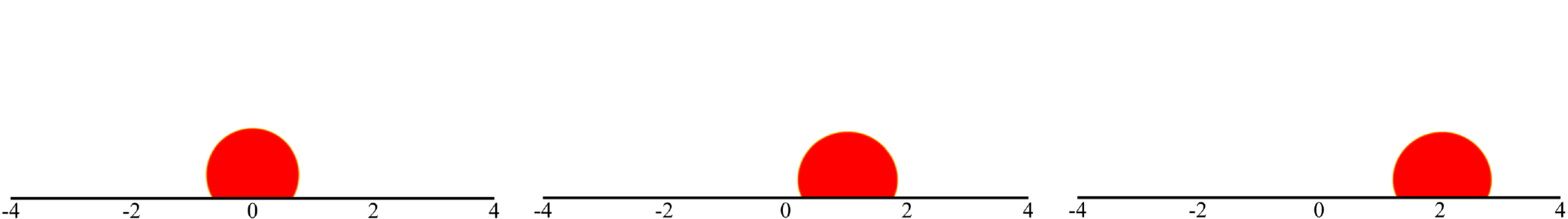}}
	\end{minipage}
	\caption{Snapshots of the displacement process for different contact angles: (a) $\theta=80^\circ$, (b) $\theta=100^\circ$, (c) $\theta=120^\circ$. In each row, images from left to right correspond to dimensionless times $ t^*=0, 4.73, 9.47$, respectively.} 
	\label{fig8}
\end{figure}

\begin{figure}[H]
	\centering
	\includegraphics[width=0.5\textwidth]{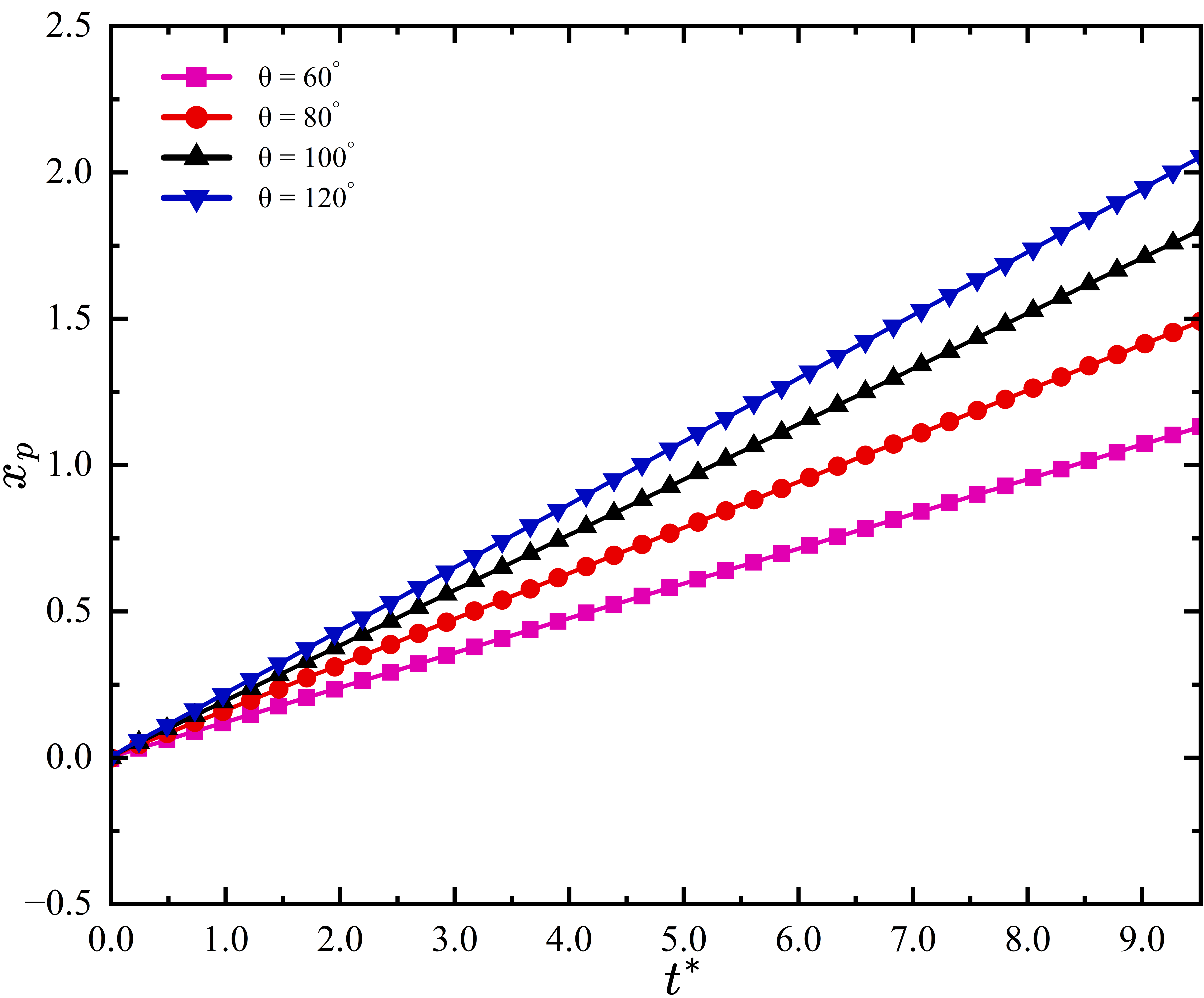}
	\caption{The relationship between centroid position ($x_p$) (normalized by the initial position) and the dimensionless time ($t^*$) at different contact angles.}
	\label{fig9}
\end{figure}

After analyzing the influence of the odd viscosity coefficient, we now focus on the effect of surface wettability on the migration behavior of odd viscous droplets.
To this end, we simulated droplet migration for four different contact angles ($60^\circ$, $80^\circ$, $100^\circ$, and $120^\circ$), while keeping the odd viscosity coefficient fixed at $\eta = 3$ and holding other physical parameters constant.
As shown in Fig. \ref{fig8}, droplets with contact angles of $80^\circ$, $100^\circ$, and $120^\circ$ all move consistently to the right.
This observation confirms that the migration direction is determined by the sign of the odd viscosity coefficient rather than by surface wettability, as demonstrated in the preceding discussion.
Additionally, Fig. \ref{fig9} presents the temporal evolution of the droplet centroid position for different contact angles.
It can be observed that the migration speed increases with the contact angle,  which is the highest at $120^\circ$ and the lowest at $60^\circ$.

\begin{figure}[H]
	\centering
	\begin{minipage}[c]{0.03\textwidth}
		\centering (a)
	\end{minipage}
	\begin{minipage}[c]{0.95\textwidth} 
		\adjustbox{valign=c}{\includegraphics[width=1\textwidth]{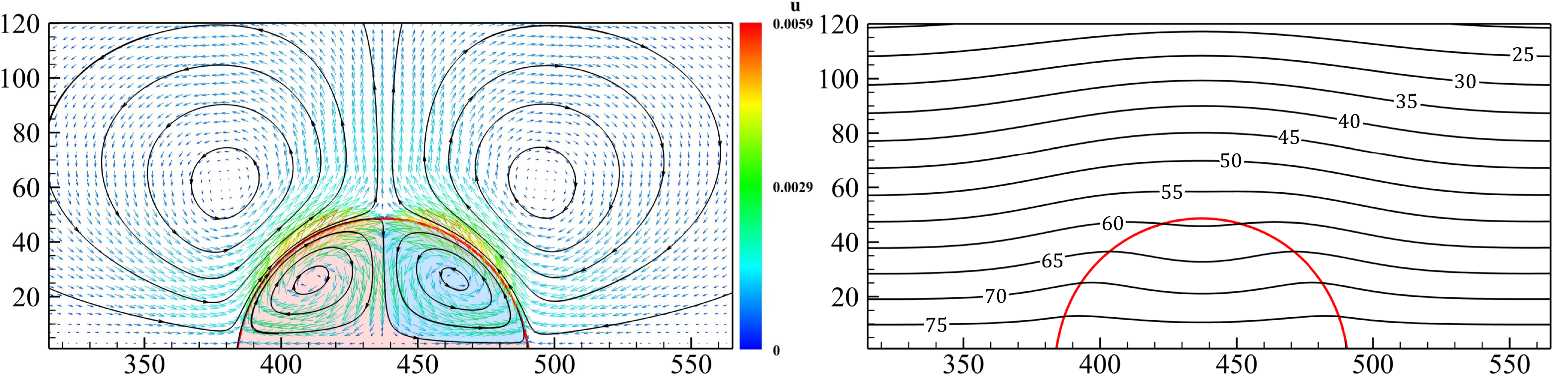}}
	\end{minipage}	
	\vspace{0.0em} 
	\begin{minipage}[c]{0.03\textwidth}
		\centering (b)
	\end{minipage}
	\begin{minipage}[c]{0.95\textwidth}
		\adjustbox{valign=c}{\includegraphics[width=1\textwidth]{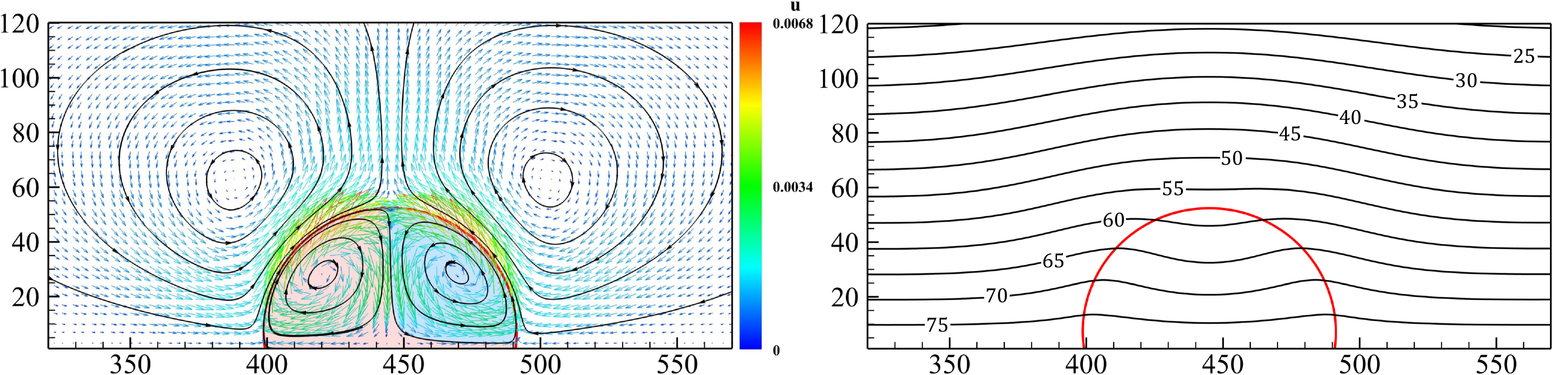}}
	\end{minipage}
	\vspace{0.0em} 
	\begin{minipage}[c]{0.03\textwidth}
		\centering (c)
	\end{minipage}
	\begin{minipage}[c]{0.95\textwidth}
		\adjustbox{valign=c}{\includegraphics[width=1\textwidth]{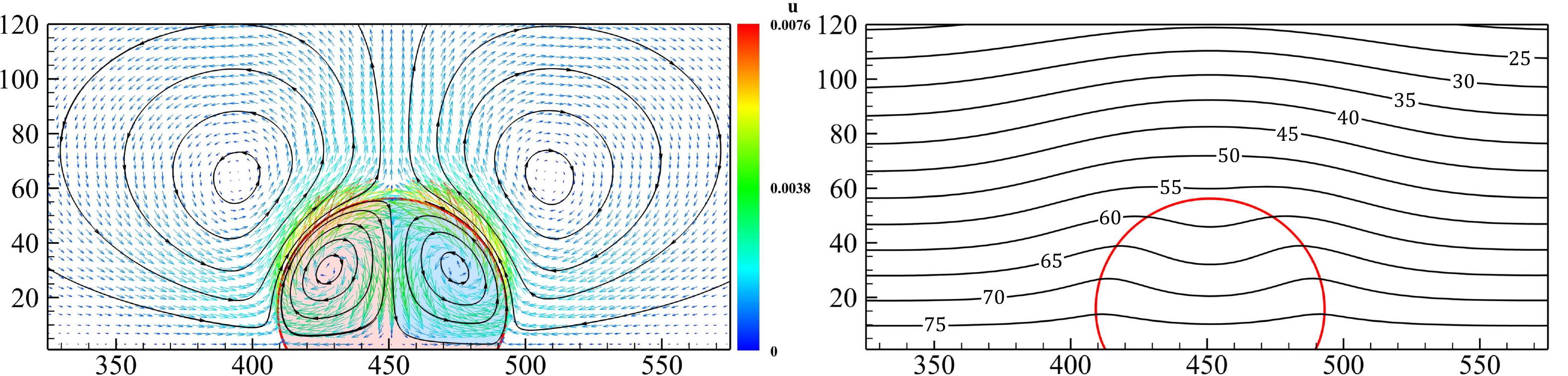}}
	\end{minipage}
	
	\caption{The flow field (the left plane) and the temperature field (the right plane) surrounding the moving droplet at different contact angles: (a) $\theta=80^\circ$, (b) $\theta=100^\circ$, (c) $\theta=120^\circ$. The red lines are $\phi = 0$, the colorful lines with arrows are the velocity vectors, and the black lines with arrows are the streamlines.} 
	\label{fig10}
\end{figure}

To gain deeper insight into how surface wettability influences the migration behavior of odd viscous droplets, we also analyze the interfacial stress response arising from the coupling between thermocapillary effects and odd viscosity.
For all contact angles, the left (right) segment of the droplet interface is always exposed to an anticlockwise (clockwise) temperature gradient.
Consequently, the surface tension gradient satisfies $\partial \gamma/\partial s > 0$ ( $\partial \gamma/\partial s < 0$) on the left (right) segment of the droplet interface.
According to Eq. \ref{eq33}, the gradient distribution results in compressive stress on the left interface and tensile stress on the right interface induced by odd viscosity. 
As shown in Fig. \ref{fig7}, such an asymmetric stress distribution consistently generates a net force that drives the droplet rightward for all contact angles. 
In addition, it is found that the migration speed increases with larger contact angles [see Fig. \ref{fig9}].
In fact, the behavior can be attributed to changes in the interfacial temperature distribution arising from variations in droplet geometry.
Considering the incompressibility condition and the absence of mass loss during migration, we approximate the droplet interface as a circular arc and assume that the droplet volume remains constant throughout the process
\begin{equation}
	\label{eq34}
	S = \frac{1}{2} \pi R^2.
\end{equation}
As a result, for a droplet with contact angle $\theta$, the radius and height are respectively given by
\begin{equation}
	\label{eq35}
	r_\theta = \sqrt{ \frac{0.5\pi}{\theta - \sin\theta \cos\theta} } R,	
	h_\theta = (1 - \cos\theta) \sqrt{ \frac{0.5\pi}{\theta - \sin\theta \cos\theta} } R,
\end{equation}
it can be known that the height of the droplet increases with the contact angle.
As seen in Figs. \ref{fig6}(c) and \ref{fig10}(a–c), a larger contact angle alters the droplet geometry, resulting in an increased droplet height that exposes the interface to a broader temperature variation.
Therefore, there exists a larger magnitude of temperature gradient $|\partial T/\partial s|$ along the droplet interface, which further  intensifies the magnitude of the surface tension gradient $|\partial \gamma/\partial s|$.
According to Eq. (\ref{eq33}), in the presence of odd viscosity, the surface tension gradient is converted into additional normal stresses via the antisymmetric stress coupling mechanism.
For the fixed $\eta$, the temperature gradientis increased as the contact angle increases. 
Then, the enhanced gradient gives rise to stronger odd compressive and tensile stresses at the left and right interfaces of the droplet, respectively, which creates an unbalanced force that ultimately accelerates the droplet's migration.

\begin{figure}[H]
	\centering
	\begin{minipage}[c]{0.03\textwidth}
		\centering(a)
	\end{minipage}
	\begin{minipage}[c]{0.95\textwidth} 
		\adjustbox{valign=c}{\includegraphics[width=1\textwidth]{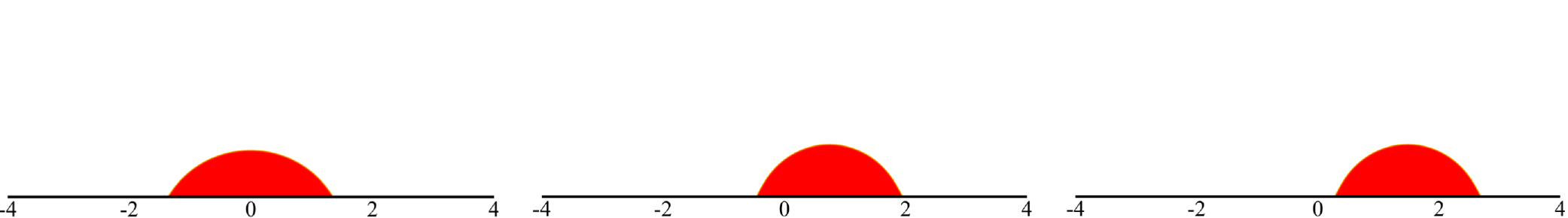}}
	\end{minipage}	
	\vspace{0.0em} 
	\begin{minipage}[c]{0.03\textwidth}
		\centering (b)
	\end{minipage}
	\begin{minipage}[c]{0.95\textwidth}
		\adjustbox{valign=c}{\includegraphics[width=1\textwidth]{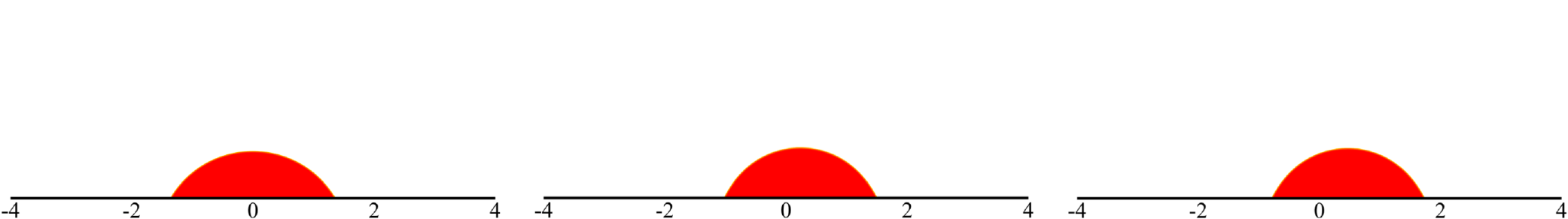}}
	\end{minipage}

	\caption{Snapshots of the displacement process for different viscosity ratios : (a) ${\nu_B}/{\nu_A} = 0.6$, (b) ${\nu_B}/{\nu_A} = 3.5$. In each row, images from left to right correspond to dimensionless times $ t^*=0, 4.73, 9.47$, respectively.} 
	\label{fig11}
\end{figure}

\begin{figure}[H]
	\centering
	\includegraphics[width=0.5\textwidth]{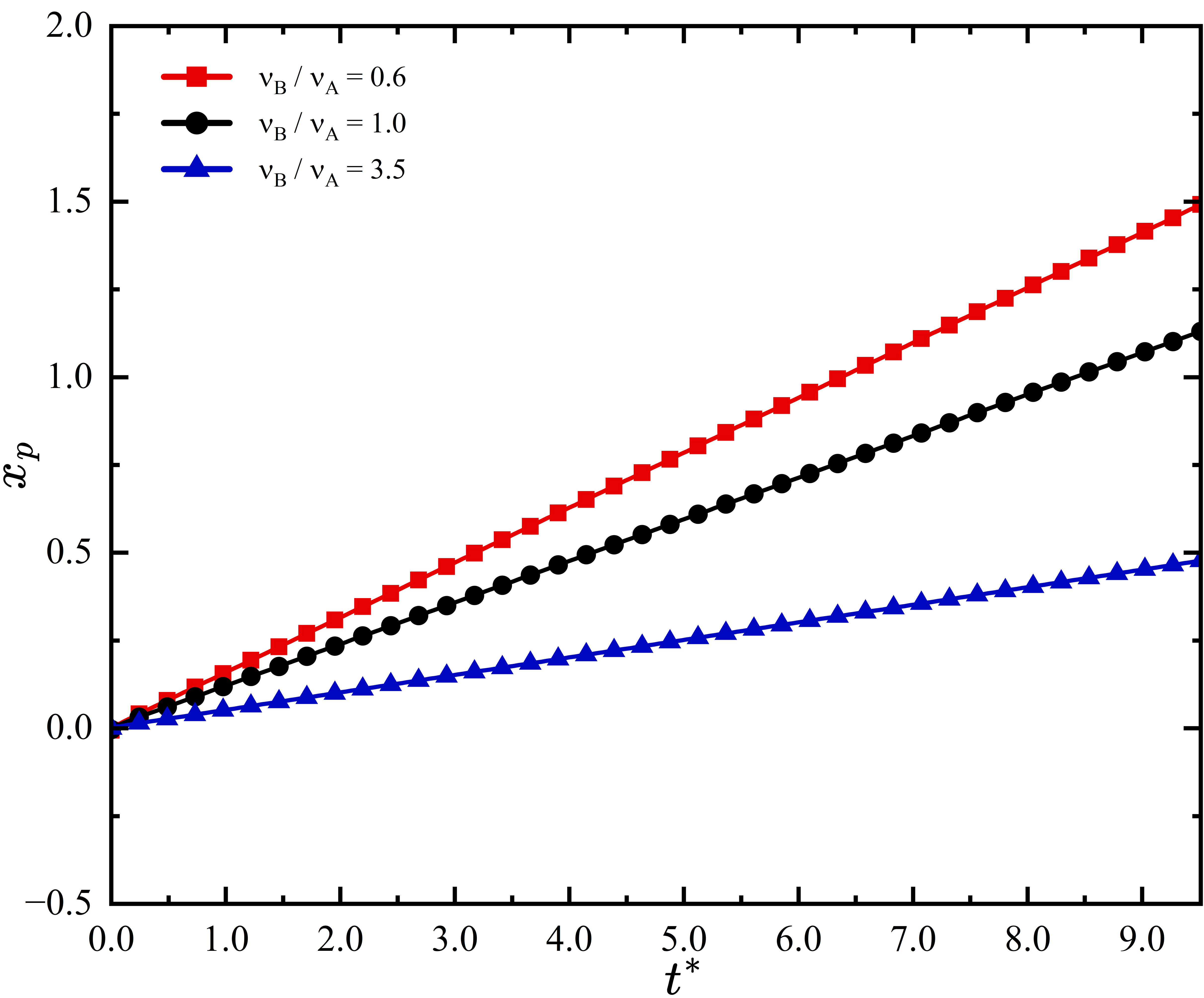}
	\caption{The relationship between centroid position ($x_p$) (normalized by the initial position) and the dimensionless time ($t^*$) at different viscosity ratios.}
	\label{fig12}
\end{figure}

\begin{figure}[H]
	\centering
	\subfigure[]{\label{fig13a}\includegraphics[width=0.49\textwidth]{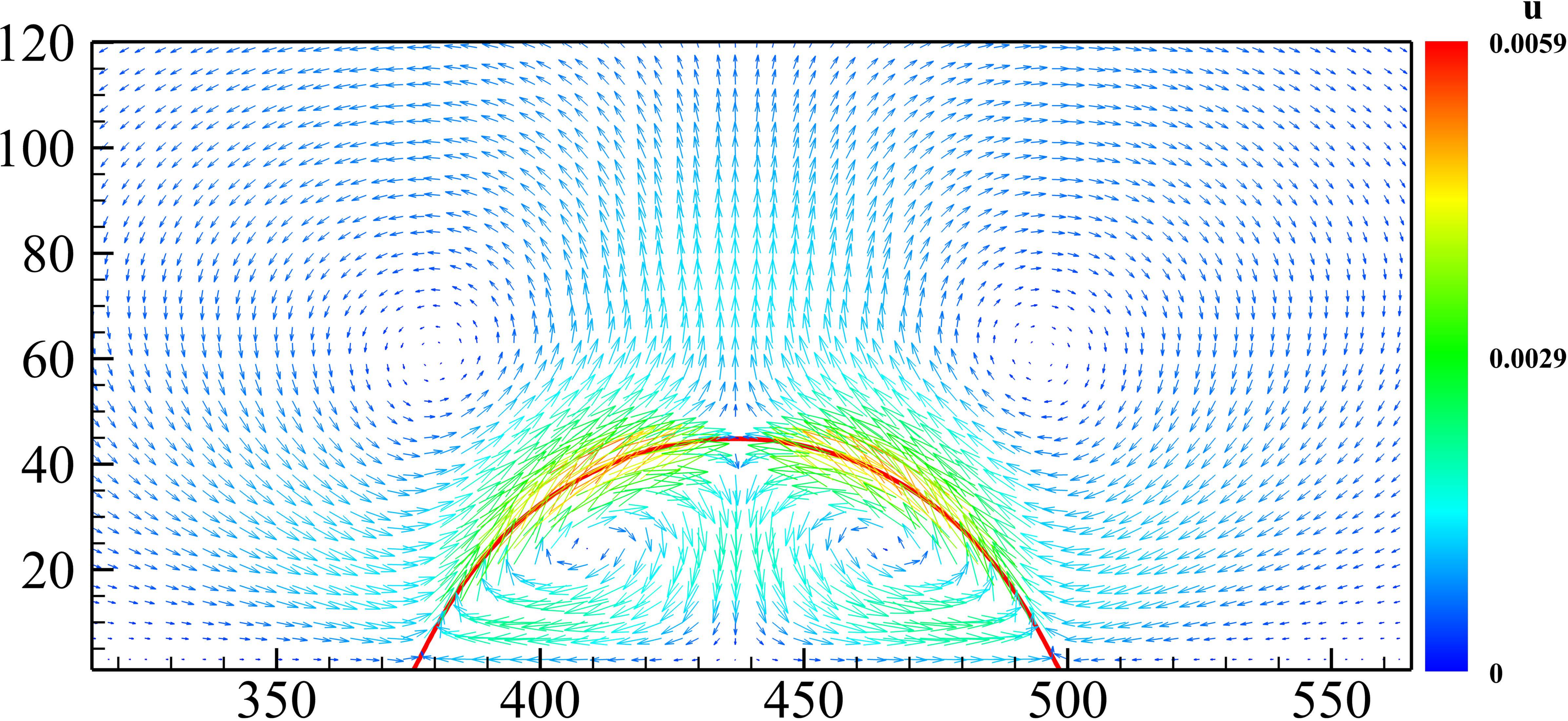}}
	\subfigure[]{\label{fig13b}\includegraphics[width=0.49\textwidth]{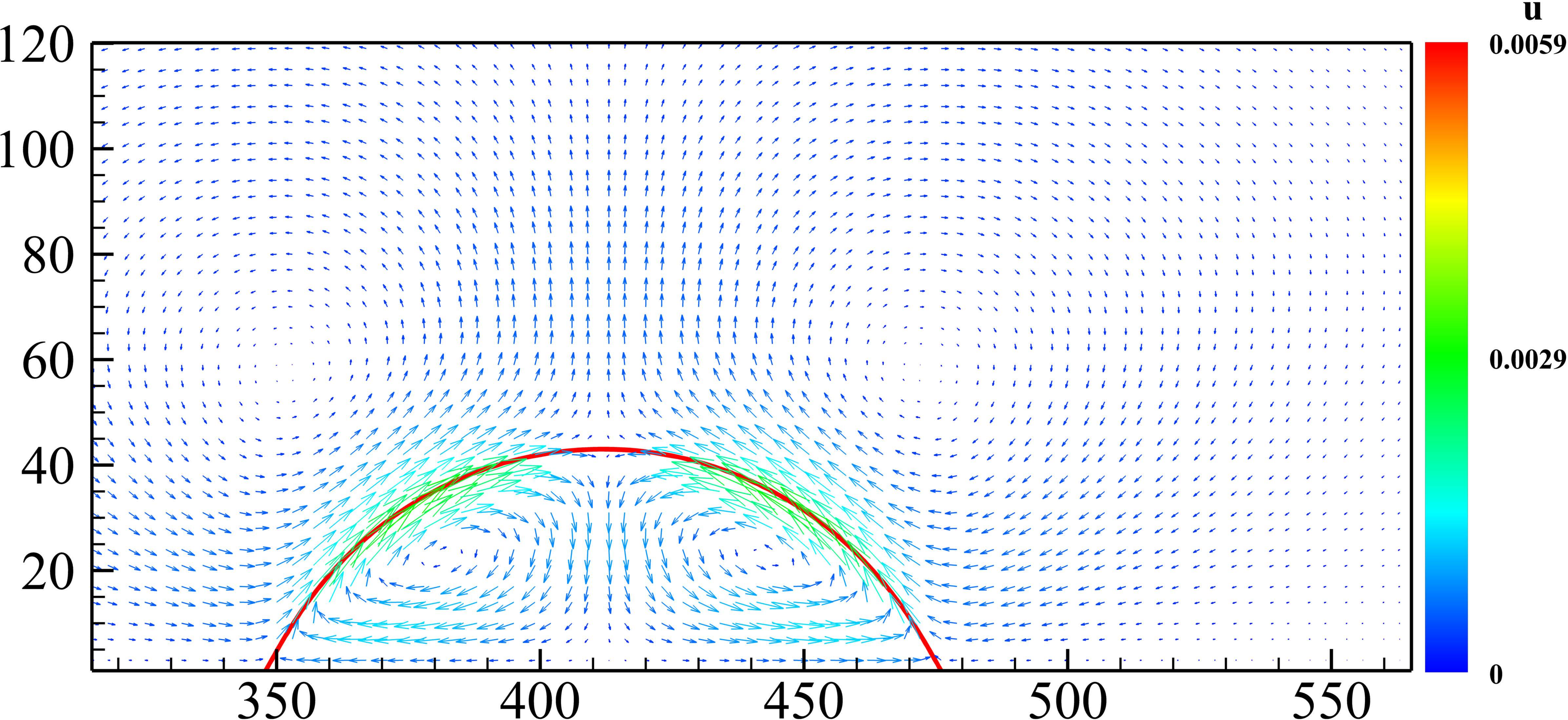}}
	\caption{The velocity ﬁeld surrounding the moving droplet at the contact angle $\theta = 60^\circ$ for different viscosity ratios: (a) ${\nu_B}/{\nu_A} = 0.6$, (b) ${\nu_B}/{\nu_A} = 3.5$, in which the red lines are $\phi = 0$, and the colorful lines with arrows are the velocity vectors.}
	\label{fig13}
\end{figure}

Having elucidated the effect of surface wettability on the migration of odd viscous droplets, we next investigate the effect of the viscosity ratio. 
To this end, we vary the viscosity of the surrounding fluid to obtain different viscosity ratios while holding the droplet viscosity constant.
All simulations are conducted at a fixed odd viscosity coefficient of $\eta = 3$, and the other parameters are kept constant.
Fig. \ref{fig11} shows the migration of a droplet on a flat substrate with a contact angle of $\theta = 60^\circ$ for viscosity ratios of $0.6$ and $3.5$. The numerical results from Fig. \ref{fig12} indicate that the droplet migration is slowed as the viscosity ratio increases. 
To understand the mechanism behind the slowdown in droplet migration with increasing viscosity ratio, the velocity field shown in Fig. \ref{fig13} is analyzed. 
As shown in this figure, it can be clearly seen that the droplet shape is almost identical to that in the case of ${\nu_B}/{\nu_A} = 1$.
In addition, the velocity field exhibits similar characteristics to the case of ${\nu_B}/{\nu_A} = 1$ [see Fig. \ref{fig6} (c)]. 
Further, another point worth being mentioned is that the vortices are found to weaken markedly as the viscosity ratio increases.
Actually, this phenomenon is  attributed to the increased viscous shear resistance on the droplet as the viscosity ratio increases, which leads to greater viscous dissipation.
As a consequence, the increased viscous dissipation reduces the net propulsion induced by odd viscosity, resulting in a slower droplet migration.

\begin{figure}[H]
	\centering
	\begin{minipage}[c]{0.03\textwidth}
		\centering(a)
	\end{minipage}
	\begin{minipage}[c]{0.95\textwidth} 
		\adjustbox{valign=c}{\includegraphics[width=1\textwidth]{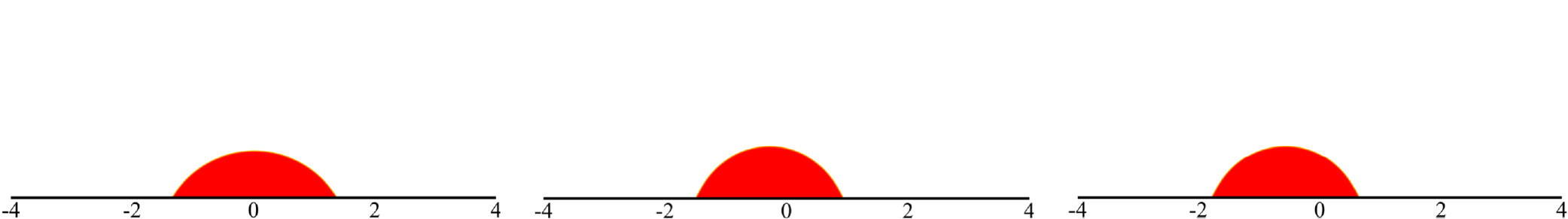}}
	\end{minipage}	
	\begin{minipage}[c]{0.03\textwidth}
		\centering (b)
	\end{minipage}
	\begin{minipage}[c]{0.95\textwidth}
		\adjustbox{valign=c}{\includegraphics[width=1\textwidth]{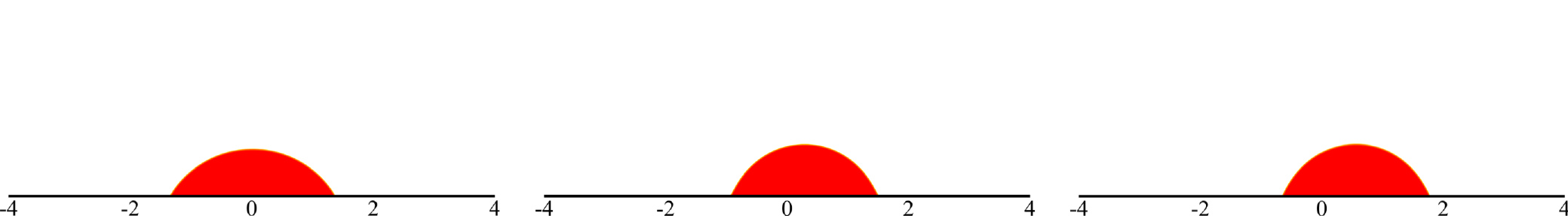}}
	\end{minipage}	
	\begin{minipage}[c]{0.03\textwidth}
		\centering (c)
	\end{minipage}
	\begin{minipage}[c]{0.95\textwidth}
		\adjustbox{valign=c}{\includegraphics[width=1\textwidth]{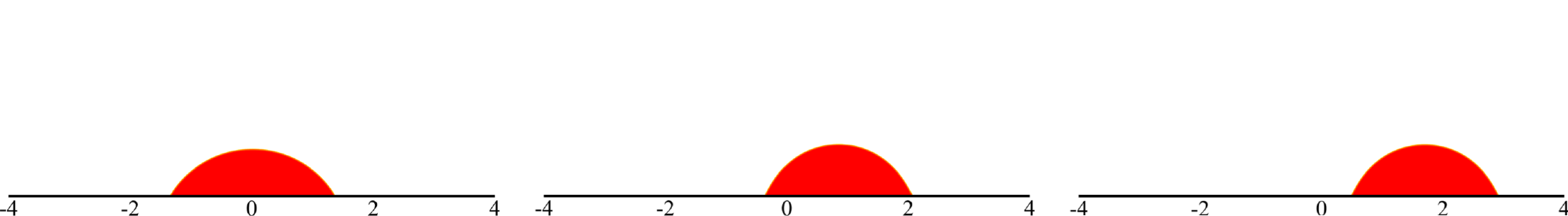}}
	\end{minipage}
	\caption{Snapshots of the displacement process at the inclination angle $\alpha=30^\circ$ for different odd viscosity coefficients: (a) $\eta =-3.0$, (b) $\eta = 0$, (c) $\eta = 3.0$. In each row, images from left to right correspond to dimensionless times $ t^*=0, 4.73, 9.47$, respectively.} 
	\label{fig14}
\end{figure}

\begin{figure}[H]
	\centering
	\subfigure[]{\label{fig15a}\includegraphics[width=0.495\textwidth]{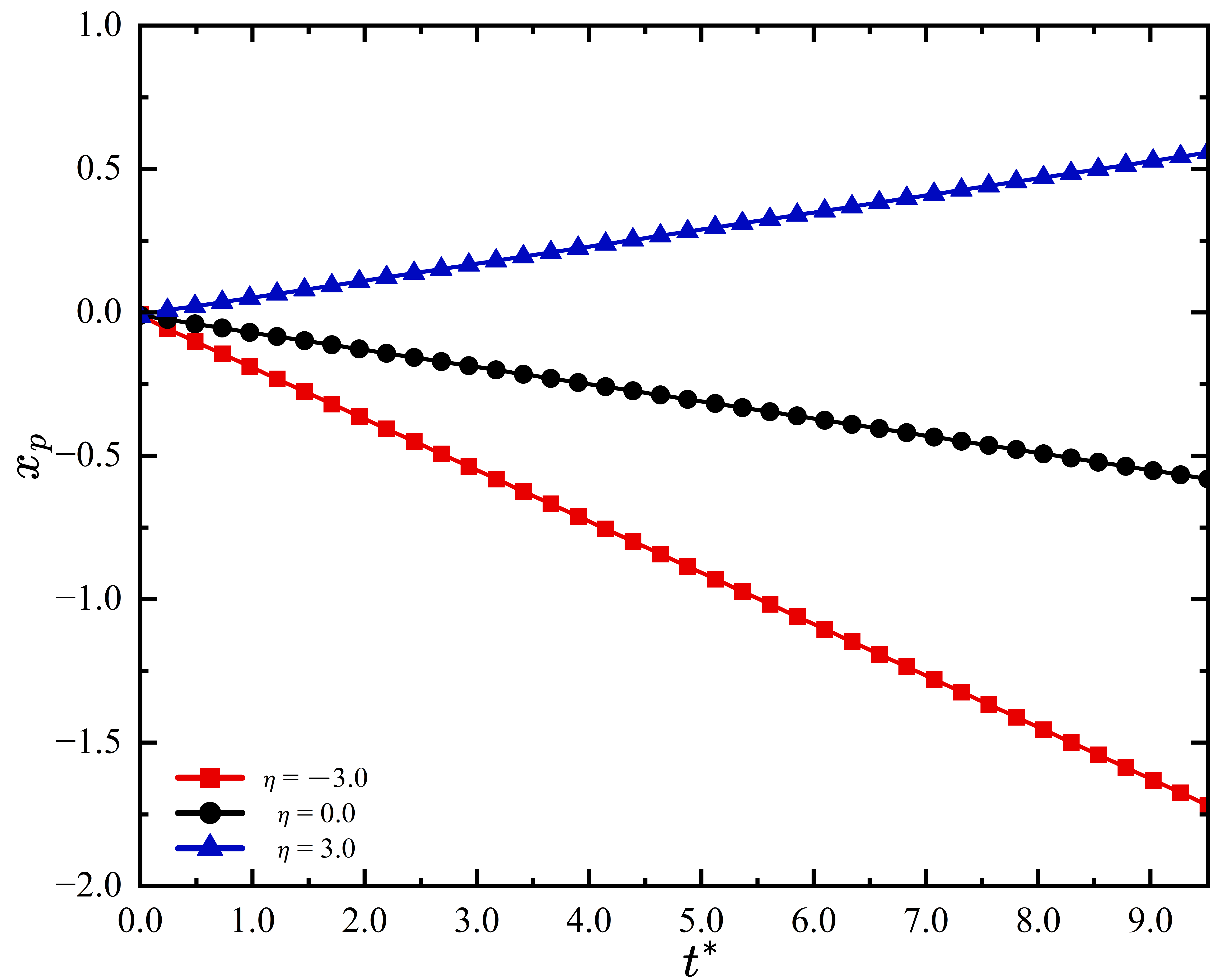}}
	\subfigure[]{\label{fig15b}\includegraphics[width=0.495\textwidth]{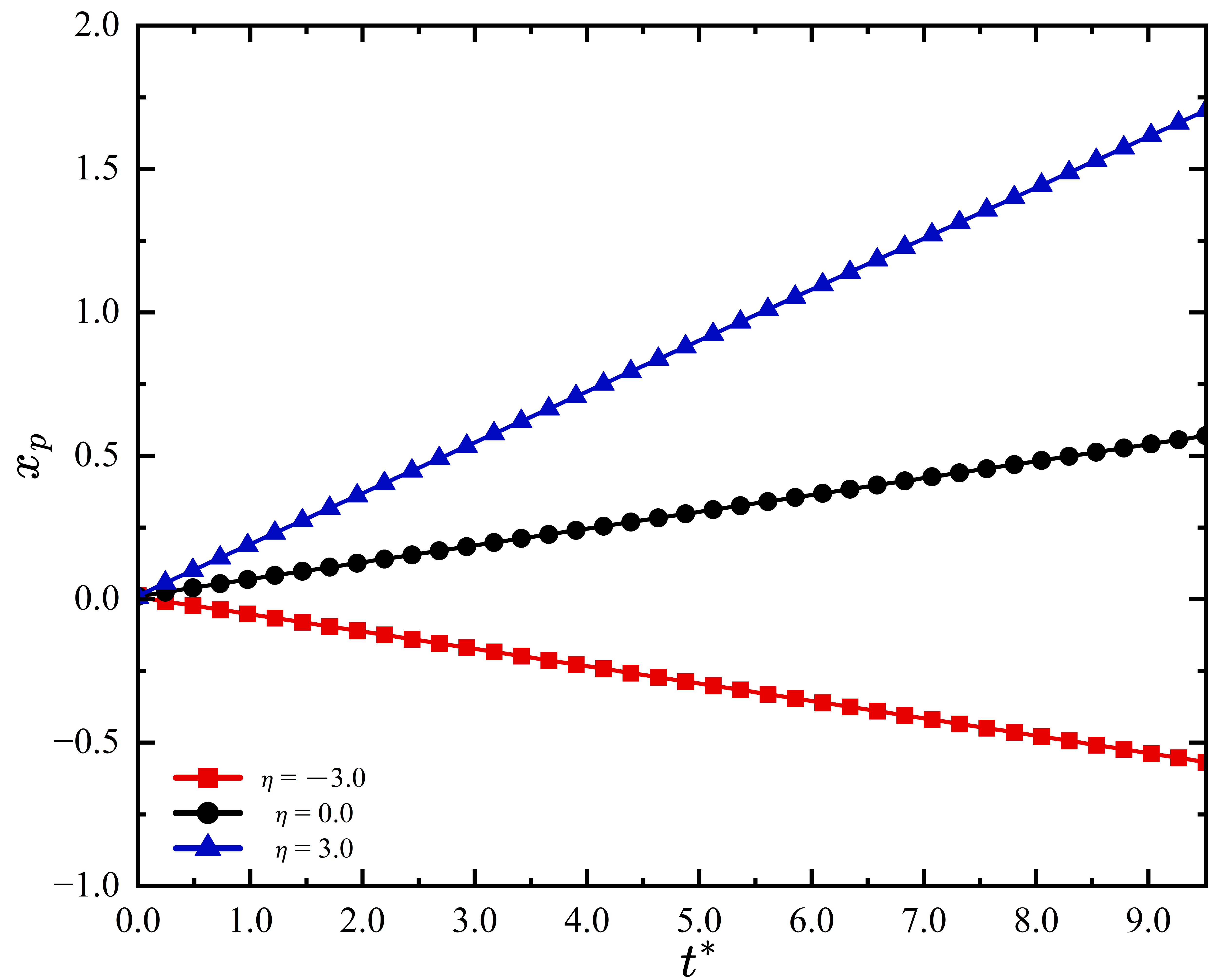}}
	\caption{The relationship between centroid position ($x_p$) (normalized by the initial position) and the dimensionless time ($t^*$) at different inclination angles: (a) $\alpha =-30^\circ$, (b) $\alpha =30^\circ$.}
	\label{fig15}
\end{figure}

The above simulations are performed on a horizontal substrate with an inclination angle of $0^\circ$. 
However, in practical applications, maintaining a perfectly horizontal surface is often difficult, and surface inclination becomes a key factor influencing droplet behavior.
To this end, the effect of surface inclination on the thermocapillary migration is further investigated.
Accordingly, two representative inclination angles are considered: $\alpha = -30^\circ$ and $\alpha=30^\circ$. All simulations are performed at a fixed contact angle of $\theta = 60^\circ$ and with odd viscosity coefficients of $\eta$ = –3, 0, and 3. 
Unlike the previous simulations on horizontal surface, the inclination angle introduces a gravitational component along the substrate, which becomes a significant factor influencing droplet migration.
Fig. \ref{fig14} shows the migration behavior of droplets at an inclination angle of $\alpha =  30^\circ$, where the gravitational component points rightward. 
For a conventional droplet ($\eta = 0$), the migration is downhill as expected [see Fig. \ref{fig14}(b)]. 
Nevertheless, the presence of odd viscosity dramatically alters the droplet's downhill motion.
Specifically, at $\eta = -3$, the droplet migrates uphill against gravity [see Fig. \ref{fig14}(a)], while at $\eta = 3$, its downhill speed increases markedly compared to $\eta = 0$ [see Fig. \ref{fig14}(c)].
Conversely, when $\alpha = -30^\circ$, the component of gravity along the substrate points to the left. 
The droplet without odd viscosity ($\eta = 0$) always migrates toward the lower end of the inclined substrate.
When $\eta = -3$, the droplet still migrates downhill, but with a noticeably higher speed. However, when $\eta = 3$, the droplet exhibits uphill migration against gravity.
A quantitative comparison is provided in Fig. \ref{fig15}, which presents the evolution of the droplet centroid position over dimensionless time for $\alpha = -30^\circ$ and $\alpha = 30^\circ$.
These plots further demonstrate that both the migration direction and speed vary significantly with $\alpha$ and $\eta$. 
Notably, in the absence of odd viscosity ($\eta = 0$), the droplet migrates downhill at identical speeds for $\alpha = \pm 30^\circ$ [see Figs. \ref{fig15}(a),(b)], 
and the centroid trajectory is symmetric with respect to the initial position, indicating that the migration behavior exhibits mirror symmetry under equal but opposite inclination angles.
It is noteworthy that the presence of odd viscosity breaks this mirror symmetry, as exemplified by the droplet with $\eta = 3$, which exhibits starkly different behaviors at opposite angles, migrating uphill at $\alpha = -30^\circ$ [see Fig. \ref{fig15}(a)], while moving downhill at $\alpha = 30^\circ$ [see Fig. \ref{fig15}(b)]. Additionally, as can be seen from the figure, the magnitude of the droplet's downhill velocity is significantly greater than that of the uphill velocity.

The migration behavior of droplets can be explained by analyzing the distribution of driving forces on the interface. 
For a conventional droplet ($\eta = 0$), gravity is the primary driver for migration. 
As shown in Figs. \ref{fig15}(a) and (b), the droplet consistently moves downhill for both $\alpha = -30^\circ$ and $\alpha = 30^\circ$. 
In contrast to the conventional case ($\eta = 0$), odd viscosity induces an antisymmetric distribution of normal stresses, generating a net driving force that competes with or assists gravity. 
The interaction between the odd viscosity-induced force and gravity is first analyzed for the case of $\alpha = -30^\circ$, where gravity points leftward [see Fig. \ref{fig15}(a)].
When $\eta < 0$, the net driving force induced by odd viscosity aligns with the gravitational component along the substrate, thereby enhancing the total driving force and resulting in increased downhill migration speed.
Conversely, when $\eta > 0$, the odd viscosity-induced force acts in the opposite direction to the gravitational component, which not only partially counteracts gravity but may also generate a net uphill driving force, leading to droplet climbing behavior.
For the opposing inclination angle of $\alpha = 30^\circ$, where gravity points rightward, odd viscosity exerts a different effect on droplet motion [see Figs. \ref{fig14}(a-c)].
Here, for $\eta < 0$, the odd viscosity-induced force acts in the opposite direction to the gravitational component, which significantly reduces the migration speed and may even lead to uphill movement.
By contrast, for $\eta > 0$, the odd viscosity-induced force now aligns with the gravitational component, and the two forces combine to accelerate the droplet’s downhill migration.

\section{Conclusions}\label{section7}
Thermocapillary migration is crucial for microscale fluid transport and has received significant attention in recent years. However, most existing studies have examined conventional droplets migrating along imposed temperature gradients. In contrast, this work explores the thermocapillary migration of an odd viscous droplet on a uniformly heated surface, and focuses on the effects of odd viscosity coefficient, surface wettability, viscosity ratio ${\nu_B}/{\nu_A}$  between the surrounding fluid and the droplet, and substrate inclination angle on the droplet’s migration behavior.

Based on the present simulation results, it is noted that the odd viscous droplet introduces an antisymmetric coupling in the stress tensor, which converts the symmetric tangential Marangoni stress into an asymmetric normal stress. This distinct feature disrupts the internal flow symmetry within the droplet, causing it to migrate along the uniformly heated substrate. 
In addition, regarding the influence of the substrate's surface wetting, it is interesting to note that an odd viscous droplet with a relatively larger contact angle exhibits a greater temperature difference within the droplet. The increased temperature difference further enhances the surface tension gradient, causing the imbalance of the normal stresses generated on both sides of the droplet by odd viscosity to become more pronounced, which in turn increases the droplet's migration speed. Furthermore, we observe that as the viscosity ratio ${\nu_B}/{\nu_A}$ increases, the interfacial viscous shear resistance also increases, leading to a reduction in the droplet's migration speed. Moreover, for an odd viscous droplet on an inclined substrate, the direction of the droplet's motion depends on the interaction between the force generated by odd viscosity and gravity. In some cases, when the force from odd viscosity surpasses gravity, it can even cause the droplet to climb upward against gravity. 

In conclusion, the findings of this study can provide a theoretical reference for designing novel microfluidic actuation and manipulation technologies. Considering the anisotropic nature of the force induced by odd viscosity droplets, the next step is to conduct a systematic study on the thermocapillary migration of three-dimensional odd viscosity droplets.

\section*{Acknowledgments}
This work is supported by the National Natural Science Foundation of China (Grant No. 12472297).


\begin{thebibliography}{1}


\bibitem{Pra Langmuir2008}V. Pratap, N. Moumen, R.S. Subramanian, Thermocapillary motion of a liquid drop on a horizontal solid surface, Langmuir 24 (9) (2008) 5185-5193.
\bibitem{Hu Langmuir2005}H. Hu, R.G. Larson, Analysis of the effects of Marangoni stresses on the microflow in an evaporating sessile droplet, Langmuir 21 (9) (2005) 3972-3980.
\bibitem{Yan Micromachines2019}Z. Yan, M. Jin, Z. Li, G. Zhou, L. Shui, Droplet-based microfluidic thermal management methods for high performance electronic devices, Micromachines 10 (2) (2019) 89.
\bibitem{Utada Science2005} A.S. Utada, E. Lorenceau, D.R. Link, P.D. Kaplan, H.A. Stone, D.A. Weitz, Monodisperse double emulsions generated from a microcapillary device, Science 308 (5721) (2005) 537-541.

\bibitem{Atencia Nature2005} J. Atencia, D.J. Beebe, Controlled microfluidic interfaces, Nature 437 (7059) (2005) 648-655.

\bibitem{Jiang Microgravity2019} H. Jiang, S. Li, L. Zhang, J. He, J. Zhao, Effect of microgravity on the solidification of aluminum-bismuth-tin immiscible alloys, npj Microgravity 5 (1) (2019) 26.

\bibitem{Young JFM1959}N.O. Young, J.S. Goldstein, M. Block, The motion of bubbles in a vertical temperature gradient, J. Fluid Mech. 6 (3) (1959) 350-356.


\bibitem{Kamotani JFM2000}Y. Kamotani, S. Ostrach, J. Masud, Microgravity experiments and analysis of oscillatory thermocapillary flows in cylindrical containers, J. Fluid Mech. 410 (2000) 211-233.

\bibitem{Kang JFM2019}Q. Kang, J. Wang, L. Duan, Y. Su, J. He, D. Wu, W. Hu, The volume ratio effect on flow patterns and transition processes of thermocapillary convection, J. Fluid Mech. 868 (2019) 560-583.

\bibitem{Brzoska Langmuir1993}J.B. Brzoska, F. Brochard-Wyart, F. Rondelez, Motions of droplets on hydrophobic model surfaces induced by thermal gradients, Langmuir 9 (8) (1993) 2220-2224.

\bibitem{Dai Langmuir2016}Q. Dai, M.M. Khonsari, C. Shen, W. Huang, X. Wang, Thermocapillary migration of liquid droplets induced by a unidirectional thermal gradient, Langmuir 32 (30) (2016) 7485-7492.

\bibitem{Baumgartner PNAS2022}D.A. Baumgartner, S. Shiri, S. Sinha, S. Karpitschka, N.J. Cira, Marangoni spreading and contracting three-component droplets on completely wetting surfaces, Proc. Natl. Acad. Sci. 119 (19) (2022) e2120432119.

\bibitem{Karbalaei Micromachines2016}A. Karbalaei, R. Kumar, H.J. Cho, Thermocapillarity in microfluidics-A review, Micromachines 7 (1) (2016) 13.

\bibitem{Bekezhanova AMM2018} V.B. Bekezhanova, O.N. Goncharova, Modeling of three dimensional thermocapillary flows with evaporation at the interface based on the solutions of a special type of the convection equations, Appl. Math. Model. 62 (2018) 145-162.

\bibitem{Li JAP2024}C. Li, X. Xie, T. Xiong, X. Ye, Thermocapillary migration of a droplet on a substrate with wettability difference: A comparison between slip and precursor film models in three dimensions, J. Appl. Phys. 135 (6) (2024).

\bibitem{Das POF2018}S. Das, S. Chakraborty, Influence of complex interfacial rheology on the thermocapillary migration of a surfactant-laden droplet in Poiseuille flow, Phys. Fluids 30 (2) (2018).

\bibitem{Wu POF2022}Z.B. Wu, Thermocapillary droplet migration in a vertical temperature gradient controlled by thermal radiations, Phys. Fluids 34 (2) (2022).

\bibitem{Nguyen EJM2024}H.D. Nguyen, T.V. Vu, N.X. Ho, P.H. Nguyen, A.D. Le, Thermocapillary migration of a compound droplet on a substrate, Eur. J. Mech. B/Fluids 103 (2024) 1-10.

\bibitem{Wang IECR2023} J.X. Wang, F.Y. Zhang, S.Y. Li, Y.P. Cheng, W.C. Yan, F. Wang, J.L. Xu, Y. Sui, Numerical studies on the controlled thermocapillary migration of a sessile droplet, Ind. Eng. Chem. Res. 62 (44) (2023) 18792-18799.

\bibitem{Kalichetty IJHMT2023}S.S. Kalichetty, T. Sundararajan, A. Pattamatta, Numerical study of thermocapillary migration of a droplet on an oleophilic track, Int. J. Heat Mass Transf. 214 (2023) 124448.

\bibitem{Yan ATE2025} H. Yan, L. Wang, J. Huang, Y. Yu, Thermocapillary migration of a self-rewetting droplet on an inclined surface: A phase-field simulation, Appl. Therm. Eng. 263 (2025) 125345.

\bibitem{Landau FM1987}L.D. Landau, E.M. Lifshitz, Fluid Mechanics: Volume 6, Elsevier, 1987.

\bibitem{Avron JSP1998}J.E. Avron, Odd Viscosity, J. Stat. Phys. 92 (3) (1998) 543-557.

\bibitem{Fruchart AR2023}M. Fruchart, C. Scheibner, V. Vitelli, Odd viscosity and odd elasticity, Annu. Rev. Condens. Matter Phys. 14 (1) (2023) 471-510.

\bibitem{Souslov PRL2019}A. Souslov, K. Dasbiswas, M. Fruchart, S. Vaikuntanathan, V. Vitelli, Topological waves in fluids with odd viscosity, Phys. Rev. Lett. 122 (12) (2019) 128001.

\bibitem{Markovich PRL2021}T. Markovich, T.C. Lubensky, Odd viscosity in active matter: Microscopic origin and 3D effects, Phys. Rev. Lett. 127 (4) (2021) 048001.

\bibitem{de Nature2024}X.M. de Wit, M. Fruchart, T. Khain, F. Toschi, V. Vitelli, Pattern formation by turbulent cascades, Nature 627 (8004) (2024) 515-521.


\bibitem{Banerjee Nature2017} D. Banerjee, A. Souslov, A.G. Abanov, V. Vitelli, Odd viscosity in chiral active fluids, Nat. Commun. 8 (1) (2017) 1573.

\bibitem{Franca 2025} H. França, M. Jalaal, Odd Droplets: Fluids with Odd Viscosity and Highly Deformable Interfaces, arXiv preprint arXiv:2503.21649, 2025.

\bibitem{Aggarwal PRL2023} A. Aggarwal, E. Kirkinis, M. Olvera de la Cruz, Thermocapillary Migrating Odd Viscous Droplets, Phys. Rev. Lett. 131 (19) (2023) 198201.

\bibitem{Sussman CP1998}M. Sussman, E. Fatemi, P. Smereka, S. Osher, An improved level set method for incompressible two-phase flows, Comput. \& Fluids 27 (5-6) (1998) 663-680.

\bibitem{Hirt JCP1981}C.W. Hirt, B.D. Nichols, Volume of Fluid (VOF) method for the dynamics of free boundaries, J. Comput. Phys. 39 (1) (1981) 201-225.

\bibitem{Graaf Langmuir2006}S. Van der Graaf, T. Nisisako, C.G.P.H. Schroën, R.G.M. Van Der Sman, R.M. Boom, Lattice Boltzmann simulations of droplet formation in a T-shaped microchannel, Langmuir 22 (9) (2006) 4144-4152.

\bibitem{Huang CF2022}J. Huang, L. Wang, K. He, Three-dimensional study of double droplets impact on a wettability-patterned surface, Comput. \& Fluids 248 (2022) 105669.

\bibitem{T.K SCS2017} T. Krüger, H. Kusumaatmaja, A. Kuzmin, O. Shardt, G. Silva, E.M. Viggen, The Lattice Boltzmann Method: Principles and Practice, Springer, 2017.

\bibitem{Xiong 2025} F. Xiong, L. Wang, J. Huang, K. Luo, A thermodynamically consistent phase-field lattice Boltzmann method for two-phase electrohydrodynamic flows, J. Sci. Comput. 103 (1) (2025) 1-32.

\bibitem{Wang PRE2023} L. Wang, K. He, H. Wang, 
Phase-field-based lattice Boltzmann model for simulating thermocapillary flows, Phys. Rev. E 108 (2023) 055306. 

\bibitem{Chiu JCP2011}P.H. Chiu, Y.T. Lin, A conservative phase field method for solving incompressible two-phase flows, J. Comput. Phys. 230 (1) (2011) 185-204.

\bibitem{Hua JCP2011}J. Hua, P. Lin, C. Liu, Q. Wang, Energy law preserving C0 finite element schemes for phase field models in two-phase flow computations, J. Comput. Phys. 230 (18) (2011) 7115-7131. 

\bibitem{Zhang AMM2022}Y. Zhang, B. Dong, X. An, Y. Wang, X. Zhou, W. Li, Phase-field-based lattice Boltzmann model for ternary fluid flows considering the wettability effect, Appl. Math. Model. 103 (2022) 195-220.

\bibitem{C.Ma IJMF2011} C. Ma, D. Bothe, Direct numerical simulation of thermocapillary flow based on the volume of fluid method, Int. J. Multiphase Flow 37 (9) (2011) 1045-1058.

\bibitem{Qian EL1992} Y.H. Qian, D. d'Humieres, P. Lallemand, Lattice BGK models for Navier-Stokes equation, Europhys. Lett. 17 (6) (1992) 479-484.

\bibitem{I.G CCP2008} I. Ginzburg, F. Verhaeghe, D. d'Humieres, Two-relaxation-time lattice Boltzmann scheme: About parametrization, velocity, pressure and mixed boundary conditions, Commun. Comput. Phys. 3 (2) (2008) 427-478.

\bibitem{Luo PRE2000}P. Lallemand, L.S. Luo, Theory of the lattice Boltzmann method: Dispersion, dissipation, isotropy, Galilean invariance, and stability, Phys. Rev. E 61 (6) (2000) 6546-6562.  

\bibitem{Wang PRE2016}H.L. Wang, Z.H. Chai, B.C. Shi, H. Liang, Comparative study of the lattice Boltzmann models for Allen-Cahn and Cahn-Hilliard equations, Phys. Rev. E 94 (3) (2016) 033304.
 
\bibitem{Ding PRE2007}H. Ding, P.D.M. Spelt, Wetting condition in diffuse interface simulations of contact line motion, Phys. Rev. E 75 (4) (2007) 046708.

\bibitem{Liu JCP2015}  H. Liu, Y. Zhang, Modelling thermocapillary migration of a microfluidic droplet on a solid surface, J. Comput. Phys. 280 (2015) 37-53.

\bibitem{Voc JCS2015} R. Vochten, G. Petre, Study of the heat of reversible adsorption at the air-solution interface. II. Experimental determination of the heat of reversible adsorption of some alcohols, J. Colloid Interface Sci. 42 (2) (1973) 320-327.


\bibitem{Oron RMP1997}A. Oron, S.H. Davis, S.G. Bankoff, Long-scale evolution of thin liquid films, Rev. Mod. Phys. 69 (3) (1997) 931-980.

\end{thebibliography}
\end{document}